\documentclass[12pt]{article}
 \usepackage[dvips]{graphicx}
 \usepackage{amsmath}
 \usepackage{amsfonts}
 \usepackage{amssymb}
 \usepackage{pst-plot}
 \usepackage{pstricks}
 \usepackage{multido}
 \usepackage{fp}
 \usepackage{ifthen}
 \usepackage{longtable}
 \usepackage{datetime2}

 \newcommand{\showlabel}[1]{%
   \label{#1}%
 }%

 \allowdisplaybreaks[4]     

 \renewcommand{\sf}{\sffamily}
 \renewcommand{\bf}{\bfseries}
 \renewcommand{\it}{\itshape}

 \renewcommand{\d}{{\rm d}}     
 
 \newcommand{\td}[2]{\frac{{\rm d} #1}{{\rm d} #2}}     
 \newcommand{\tdil}[2]{{\rm d} #1/{\rm d} #2}     
 \newcommand{\pd}[2]{\frac{\partial #1}{\partial #2} }     

 \newcommand{\ds}{\displaystyle}

 \newcommand{\mb}{\mathbf}

 \newcommand{\be}{\mbox{\bf \rm e}}

 \newcommand{\ol}[1]{\overline{#1}}

 \newcommand{\la}{\leftarrow}

 \newcommand{\nn}{\nonumber}
 \newcommand{\er}[1]{(\ref{#1})}          

 \renewcommand{\L}{Lema\^{\i}tre}
 \newcommand{\LT}{{\L}-Tolman}
 \newcommand{\Rt}{\dot{R}}
 
 \newcommand{\E}{\mathbb E}
 \newcommand{\cD}{{\mathcal D}}

 \newcommand{\pb}[2]{                       
    \parbox[t]{#1}{
       \raggedright
       \setlength{\parskip}{1.2ex}          
       #2
    }
 }

 \definecolor{DGreen}{rgb}{0,0.8,0}
 \definecolor{Purple}{rgb}{0.45,0,0.45}

\begin{document}
 \sffamily

 \title{Rotation, Embedding and Topology for the Szekeres Geometry}

 \author{
  Charles Hellaby
  \thanks{\tt Charles.Hellaby@uct.ac.za} \\
  {\small \it Dept. of Maths. and Applied Maths,
  University of Cape Town,
  Rondebosch,
  7701,
  South Africa}
  \and
  Robert G. Buckley
  \thanks{\tt Robert.Buckley@utsa.edu} \\
  {\small \it Dept. of Physics and Astronomy,
  University of Texas at San Antonio,
  San Antonio,
  Texas 78249,
  USA}
 }

 \date{}


 \maketitle

 \begin{abstract}
   Recent work on the Szekeres inhomogeneous cosmological models uncovered a surprising rotation effect.  Hellaby showed that the angular $(\theta, \phi)$ coordinates do not have a constant orientation, while Buckley and Schlegel provided explicit expressions for the rate of rotation from shell to shell, as well as the rate of tilt when the 3-space is embedded in a flat 4-d Euclidean space.  We here investigate some properties of this embedding, for the quasi-spherical recollapsing case, and use it to show that the two sets of results are in complete agreement.  We also show how to construct Szekeres models that are closed in the `radial' direction, and hence have a `natural' embedded torus topology.  Several explicit models illustrate the embedding as well as the shell rotation and tilt effects.
 \end{abstract}

 \section{Previous Work}

 Since its discovery, the Szekeres inhomogeneous cosmological model has always intrigued relativists, having no Killing vectors on the one hand, and yet still being silent on the other hand.  However, the study of this metric has been limited by its relative complication, and, in the cases of the planar and hyperboloidal models, with $\epsilon \ne +1$, lack of a Newtonian analogy from which to derive physical understanding.  Still, precisely because it is one of the most realistic inhomogeneous exact solutions of Einstein's field equations, which gives it much potential for application in modelling relatively complex cosmological structures on a range of scales, a fuller description of the geometry and the evolution of this spacetime is indispensable to a proper physical understanding.

 Szekeres models can be viewed as distortions of {\LT} and Ellis models.  The three arbitrary functions of the {\LT}-Ellis metrics appear in the Szekeres metric playing essentially the same physical roles, but there are three more arbitrary functions that control the deviation from spherical, planar, or pseudo-spherical (hyperboloidal) symmetry.  All 6 arbitrary functions depend on the coordinate $r$, which is a sort of `radial' coordinate label for comoving 3-surfaces in the quasi-spherical case, and, in the quasi-planar and quasi-hyperboloidal cases, whatever the appropriate equivalent happens to be.

 It was obvious from the beginning that the constant $r$ 2-surfaces (on each time slice) were not arranged `concentrically', and in Szekeres' original paper \cite{Szek75a} he described the 3-spaces as ``a set of displaced or `non-concentric' spheres, planes and pseudospheres.''

 This non-concentric arrangement was understood and taken into account when slices through Szekeres models were plotted, for example
 \cite{Bole06,Bole07,BolSus11,KraBol12a,SusGas15,SuGaHi15,BolNazWil16,Bole17,GaHiSuQu18}.
 Again, because of the relative complication of doing this, it hasn't been attempted very often.  Since the native Szekeres coordinates are stereographic, the calculation of slices typically involved converting to angular $r$-$\theta$-$\phi$ coordinates, and then projecting into Cartesian-style $X$-$Y$-$Z$ coordinates.  However, this method depends on the unstated assumption that the $\theta$-$\phi$ coordinates maintain a `constant' orientation in some sense.

 So it came as a considerable surprise when it was discovered, only recently, that, in the $\theta$-$\phi$ representation, the constant $r$ shells are also rotated relative to each other, except in highly specialised cases.  In 2013, Buckley \& Schlegel \cite{BucSch13} stated that the constant $r$ shells are rotated relative to each other by specific amounts, but did not explain how this was obtained.  Unaware of this, Hellaby, in 2017 \cite{Hell17}, questioned whether the $\theta$ \& $\phi$ coordinates represent a constant orientation, and by studying the variation of a suitable orthonormal tetrad, showed that they don't.  We refer to this paper as ``FR''.  Two years later, Buckley \& Schlegel \cite{BucSch19} provided rigorous support for their relative rotation of adjacent shells by developing a local embedding, which required not only the previously stated rotations, but also higher dimensional rotations (or `tilts') in the embedding space.  In particular, it demonstrated that these rotations straighten up geodesics.  Clearly, this effect must be incorporated into graphical routines for generating slices through Szekeres models, and methods for doing this were presented.  We refer to this paper as ``PG''.

 Confirming that the PG shell rotation and embedding result does indeed explain the origin of the frame rotations found in FR, would be a confirmation of both papers, and a useful validation of the new understanding of the Szekeres geometry, as well as the correct way to plot it graphically.

 Further, the consideration of embeddings opens up the question of whether less obvious topologies are possible.

 \section{Background}

 \subsection{The \LT\ Spacetime}

 The \LT\ (LT) spacetime \cite{Lem33,Tol34} represents a spherically symmetric cloud of dust particles that is inhomogeneous in the radial direction; both the density and the rate of expansion or contraction can vary with radius.  Its metric is
 \begin{align}
   \d s^2 = - \d t^2 + \frac{R'^2\, \d r^2}{1 + f} + R^2 \big( \d \theta^2 + \sin^2\theta \d\phi^2 \big) ~,
 \end{align}
 where $R = R(t,r)$ is the areal radius, and $f = f(r)$ is a geometry-energy factor.  From the Einstein field equations (EFEs), the evolution equation and the density are
 \begin{align}
   \Rt^2 & = \frac{2 M}{R} + f + \frac{\Lambda R^2}{3}   \showlabel{RtDE} ~, \\
   \kappa \rho & = \frac{2 M'}{R^2 R'} ~,
 \end{align}
 where $M = M(r)$ is the total gravitational mass interior to each constant $r$ shell, and $\Lambda$ is the cosmological constant.  When $\Lambda = 0$, the evolution equation \er{RtDE} has parametric solutions, for example when $f < 0$,
 \begin{align}
   R = \frac{M}{(- f)} \big( 1 - \cos\eta \big) ~,~~~~~~ t = a + \frac{M}{(- f)^{3/2}} \big( \eta - \sin\eta \big) ~.
   \showlabel{RptpEll}
 \end{align}
 and $a = a(r)$ is the local `bang time'; the time on each constant $r$ worldline at which $R(t, r) = 0$ is $t = a$.  The factors
 \begin{align}
   L = \frac{M}{(- f)} ~~~~\mbox{and}~~~~ T = \frac{M}{(- f)^{3/2}}
   \showlabel{ScLT}
 \end{align}
 can be thought of as a scale length and a scale time for the worldline at $r$; $2 L$ is the maximum areal radius reached, and $2 \pi T$ is the duration from bang to crunch.  The derivative $R'$ that appears in the metric can be expressed parametrically as
 \begin{align}
   R' & = \frac{M'}{(- f)} (1 - \phi_1) (1 - \cos\eta)
         - \frac{f' M}{f^2} \left( \frac{3}{2} \phi_1 - 1 \right) (1 - \cos\eta) \nn \\
      &~~~~ - (-f)^{1/2} a' \phi_2 (1 - \cos\eta) ~,
         \showlabel{RrParam} \\
   \mbox{where}~~~~~~~~ & \phi_1 = \frac{\sin\eta (\eta - \sin\eta)}{(1 - \cos\eta)^2} ~,~~~~~~ \phi_2 = \frac{\sin\eta}{(1 - \cos\eta)^2} ~.
 \end{align}

 The Ellis \cite{Elli67} metrics are equivalents of the LT case, having planar and hyperboloidal (pseudo-spherical) symmetry.

 \subsection{The Szekeres Spacetime}

 The Szekeres (S) spacetimes \cite{Szek75a,Szek75b} can be thought of as distortions of the LT and Ellis ones.  In addition to the free functions $f(r)$, $M(r)$ and $a(r)$ they have 3 more functions $S(r)$, $P(r)$ and $Q(r)$ that specify the deviation from symmetry --- spherical, planar, or hyperboloidal.  The metric is
 \begin{align}
   \d s^2 = - \d t^2 + \frac{\left( R' - \dfrac{R E'}{E} \right)^2 \d r^2}{\epsilon + f}
      + \frac{R^2}{E^2} \big( \d p^2 + \d q^2 \big) ~,
      \showlabel{ds2Sz}
 \end{align}
 where $\epsilon = +1, 0, -1$, and
 \begin{align}
   E = \frac{S}{2} \left( \frac{(p - P)^2}{S^2} + \frac{(q - Q)^2}{S^2} + \epsilon \right) ~.
 \end{align}
 Since $S = 0$ is not possible for a regular metric, we assume $S > 0$.  In fact the last term in \er{ds2Sz} is $R^2$ times the 2-metric for a unit sphere, pseudo-sphere, or plane, depending on whether $\epsilon$ is $+ 1$, $- 1$, or $0$.  Thus the 3-spaces are foliated by a collection of symmetric 2-spaces, but they are not {\em arranged \/} symmetrically, as we shall see.  By the EFEs, $R(t, r)$ obeys exactly the same evolution equation \er{RtDE}, while the density $\rho$ is more complicated,
 \begin{align}
   \kappa \rho & = \frac{2 \left( M' - \dfrac{3 M E'}{E} \right)}{R^2 \left( R' - \dfrac{R E'}{E} \right)} ~.
   \showlabel{rhoSz}
 \end{align}
 For more information about the Szekeres metric, see for example \cite{Kras97,HelKra02,HelKra08,Hell09,WalHel12,BucSch13,BucSch19}.

 \subsection{Angular Form of the Szekeres Metric}

 The standard stereographic mapping, for the $\epsilon = +1$ case,
 \begin{align}
   p = P + S \cot\left(\frac{\theta}{2}\right) \cos\phi ~,~~~~~~
   q = Q + S \cot\left(\frac{\theta}{2}\right) \sin\phi ~,
 \end{align}
 transforms the metric into a more complicated, non-diagonal form,
 \begin{align}
   \d s^2 & = \Bigg[ \frac{1}{\epsilon + f} \left(
         R' + \frac{R}{S} \big\{ S' \cos\theta + \sin\theta \big( P' \cos\phi + Q' \sin\phi \big) \big\} \right)^2 \nn \\
      &~~~~ + \frac{R^2}{S^2} \big\{ S' \sin\theta + (1 - \cos\theta) \big( P' \cos\phi + Q' \sin\phi \big) \big\} \nn \\
      &~~~~ + \frac{R^2}{S^2} \big\{ (1 - \cos\theta)^2 \big( P' \sin\phi - Q' \cos\phi \big) \big\} \Bigg] \d r^2 \nn \\
      &~~~~ - \frac{R^2}{S} \big\{ S' \sin\theta + (1 - \cos\theta) \big( P' \cos\phi + Q' \sin\phi \big) \big\}
         \, \d r \, \d\theta \nn \\
      &~~~~ - \frac{R^2 \sin\theta}{S} \big\{ (1 - \cos\theta)^2 \big( P' \sin\phi - Q' \cos\phi \big) \big\}
         \, \d r \, \d\phi \nn \\
      &~~~~ + R^2 (\d\theta^2 + \sin^2\theta \, \d\phi^2) ~.
   \showlabel{ds2SzAng}
 \end{align}
 For each $(t, r)$ 2-sphere, it is sometimes convenient to define a {\em local\/} cartesian frame by
 \begin{align}
 \begin{aligned}
   x & = R \sin\theta \cos\phi \\
   y & = R \sin\theta \sin\phi \\
   z & = R \cos\theta ~.
 \end{aligned}
   \showlabel{LocCartFr}
 \end{align}

 \subsection{The Szekeres Dipole}

 The factor $\cD = E'/E$ that appears in both the metric \er{ds2Sz} and the density \er{rhoSz} controls the deviation from spherical, hyperboloidal or planar symmetry, and for $\epsilon \neq 0$ it behaves like a dipole.  The dipole has maximum value and orientation
 \begin{align}
 \begin{aligned}
   \cD_m = \left. \frac{E'}{E} \right|_m & = \frac{J}{S} ~,~~~~~~
      p_m - P = \frac{P' S \big\{ \epsilon S' - J \big\}}{H^2} ~,~~~~~~
      q_m - Q = \frac{Q' S \big\{ \epsilon S' - J \big\}}{H^2} ~, \\
   \mbox{where} & ~~~~~~~~ J = \sqrt{\epsilon^2 S'^2 + \epsilon \big( P'^2 + Q'^2 \big)}\; ~,~~~~~~~
      H = \sqrt{P'^2 + Q'^2}\; ~.
 \end{aligned}
   \showlabel{SzDplpq}
 \end{align}
 The locus $E' = 0$ lies on the $(p,q)$ circle 
 \begin{align}
   \left( \frac{(p - P) S'}{S} + P' \right) + \left( \frac{(q - Q) S'}{S} + Q' \right) = P'^2 + Q'^2 + \epsilon S'^2 ~.
 \end{align}
 For the quasi-spherical case, $\epsilon = +1$, the dipole function can be written as
 \begin{align}
   \cD = \frac{E'}{E} = - \frac{S' \cos\theta + \sin\theta \big( P' \cos\phi + Q' \sin\phi \big)}{S} ~,
 \end{align}
 and it is evident that $E'/E$ ranges between opposite extremes, passing through zero on an `equatorial' circle.  Note that here $E'$ still represents the $r$ derivative at constant $p$ \& $q$.  The angular position of the dipole maximum is found from
 \begin{align}
   \sin\theta_m & = \frac{- H}{J} ~,~~~~~~
      \cos\theta_m = \frac{- S'}{J} ~,~~~~~~
      \cos\phi_m = \frac{P'}{H} ~,~~~~~~
      \sin\phi_m = \frac{Q'}{H} ~,
 \end{align}
 or, expressed in local Cartesian coordinates, the position of the maximum on the $(x, y, z)$ unit sphere is
 \begin{align}
   x_m & = \sin\theta_m \cos\phi_m = \frac{- P'}{J} ~,~~~~~~
      y_m = \sin\theta_m \sin\phi_m = \frac{- Q'}{J} ~,~~~~~~
      z_m = \cos\theta_m = \frac{- S'}{J} ~,
      \showlabel{xyzm}
 \end{align}
 while the $E' = 0$ locus is in the plane
 \begin{align}
   P' x + Q' y + S' z = 0 ~.
   \showlabel{Er0xyz}
 \end{align}
 The dipole has two obvious effects --- in the $g_{rr}$ component of \er{ds2Sz} it creates a non-uniform separation between adjacent 2-spheres of constant $r$, and in \er{rhoSz}, it creates a variation of the density distribution around each sphere.  These effects are in addition to the $r$-dependent inhomogeneity of the underlying LT model.

 \subsection{The Rotations}
 \showlabel{SzRot}

 Relative to the angular form of the Szekeres metric \er{ds2SzAng}, there is another more subtle effect of $S$, $P$ \& $Q$.  The angular coordinates $\theta$ \& $\phi$ of \er{ds2SzAng} do not in fact represent a constant orientation, and their cardinal directions do not parallel transport from one shell to the next.  Adjacent 2-spheres have a relative rotation: the sphere at $r + \delta r$ is rotated
 \begin{subequations}
 \begin{align}
   \mbox{by}~~~~ & \frac{Q'}{S} \delta r ~~~~\mbox{about the $x$ axis} \\
   \mbox{and by}~~~~ & \frac{- P'}{S} \delta r ~~~~\mbox{about the $y$ axis} ~,
 \end{align}
   \showlabel{ShellRot}%
 \end{subequations}
relative to the one at $r$ \cite{BucSch13}.  Four justifications for this were given in \cite{BucSch19}.  Firstly there is an argument about nearest points on the two spheres, explained in PG, section V.B and fig 4.  Secondly, it was shown that geodesics look much straighter once these rotations are incorporated into plots.  Thirdly was the calculation in PG appendix C, perhaps not entirely rigorous, that added displacements and rotations to the LT metric, ending up with the S metric.  Fourthly, an embedding of any given constant $t$ 3-space of the angular S metric for $\epsilon = +1$, into a 4-d space that is flat or has constant curvature, turned out to require the above-stated rotations.

 \section{Embeddings}

 \subsection{Global Embedding of Positively Curved LT 3-Spaces}

 Let $\E^4$ be a 4-d Euclidean space with Cartesian coordinates, $X, Y, Z, W$, and let the LT 3-spaces have positive curvature, $-1 \le f \le 0$.  At a fixed time $t$, $R$ becomes a function of $r$ only.  We define a 3-surface $\Sigma$ by
 \begin{subequations}
 \begin{align}
   X & = R(r) \sin\theta \cos\phi ~, \\
   Y & = R(r) \sin\theta \sin\phi ~, \\
   Z & = R(r) \cos\theta ~, \\
   W & = \int_0^r R'(r) \alpha(r) \, \d r ~,
 \end{align}
   \showlabel{LTXYZW}%
 \end{subequations}
so then
 \begin{subequations}
 \begin{align}
   \d X & = R' \sin\theta \cos\phi \, \d r + R \cos\theta \cos\phi \, \d\theta - R \sin\theta \sin \phi \, \d\phi ~, \\
   \d Y & = R' \sin\theta \sin\phi \, \d r + R \cos\theta \sin\phi \, \d\theta + R \sin\theta \cos \phi \, \d\phi ~, \\
   \d Z & = R' \cos\theta \, \d r - R \sin\theta \, \d\theta ~, \\
   \d W & = R' \alpha(r) ~.
 \end{align}
   \showlabel{LTdXYZW}%
 \end{subequations}
Here $R(r)$ is the LT areal radius $R(t,r)$ at a particular time, and $\alpha$ is
 \begin{align}
   \alpha & = \pm \sqrt{\frac{- f(r)}{1 + f(r)}}\; ~.
   \showlabel{alphaDef}
 \end{align}
 Now in \cite{Hell87} the function $f$ was interpreted as $f = - \cos^2\psi$ where $\psi$ is the angle of the tangent cone to the embedded surface at $r$, and in this notation, then, $\alpha = \cot\psi$:
 \begin{align}
   \tan\psi = \td{R}{W} = \frac{R'}{\left| R' \sqrt{\dfrac{- f}{1 + f}}\; \right|} = \pm \sqrt{\dfrac{1 + f}{- f}}\; = \frac{1}{\alpha} \\
   \to~~~~~~~~ \cos\psi = \sqrt{- f}\; ~,~~~~ \sin\psi = \sqrt{1 + f}\; ~,~~~~ \cot\psi = \alpha ~.
 \end{align}
 Since $f \le 0$, closed models are quite likely, in which case there will be at least one point $r_m$ that is a maximum (or minimum) in $R$ where $R' = 0$, $f = - 1$, but $R'/\sqrt{1 + f}\;$ is finite \cite{HelLak85,Hell87}.  As a spatial extremum is approached and traversed, $R'$, $\psi$ and $\alpha$ change sign, $R'$ \& $\psi$ passing through zero and $\alpha$ diverging.  This ensures $\tdil{W}{r}$ retains a constant sign%
 \footnote{\sf This is not essential, and other choices could lead to different valid embeddings.}%
 .

 Using this embedding, \er{LTXYZW} and \er{LTdXYZW} show that the metric of the 3-surface becomes
 \begin{align}
   \d X^2 & + \d Y^2 + \d Z^2 + \d W^2 =
      \frac{R'^2 \, \d r^2}{1 + f} + R^2 \big( \d\theta^2 + \sin^2\theta \, \d\phi^2 \big)
 \end{align}
 which is the spatial part of the LT metric, $\d t = 0$.

 In the case that
 \begin{align}
   R = K \sin r ~,~~~~ R' = K \cos r & ~,~~~~ f = - \sin^2 r ~,~~~~ K ~\mbox{constant,}
 \intertext{we find}
   \frac{R'^2 \, \d r^2}{1 + f} + R^2 \big( \d\theta^2 + \sin^2\theta \, \d\phi^2 \big)
      & = K^2 \Big( \d r^2 + \sin^2 r \big( \d\theta^2 + \sin^2\theta \, \d\phi^2 \big) \Big)
 \end{align}
 and $\Sigma$ becomes the 3-sphere.

 \subsection{Buckley \& Schlegel's Local Szekeres Embedding}
 \showlabel{BEmbed}

 For quasi-spherical S models with $f < 0$, their constant $t$ 3-spaces can be embedded in a 4-d flat space, but for general $f$, one must embed in a 4-space of constant curvature \cite{BucSch19}.  We here consider the former case.  Let $\E^4$ be a 4-d Euclidean space with Cartesian coordinates, $X, Y, Z, W$.  In this 4-space, a 3-surface is constructed from a sequence of 2-spheres, by expanding, shifting and rotating a unit sphere, as a function of parameter $r$.  Suitable choices of the expansion, shift and rotation functions ensure the intrinsic metric of the 3-surface is identical with the positively curved, quasi-spherical Szekeres 3-spaces of constant $t$.  While an embedding is primarily a visualisation tool, in this case it also provides a clear confirmation of Buckley and Schlegel's rotations, given in equation \er{ShellRot}.

 It is also convenient to define local Cartesian coordinates, $x, y, z, w$ near each constant $r$ shell.  In these coordinates, the 2-spherical shell lies in the $w = 0$ 3-space, according to \er{LocCartFr}, so the accumulated displacements and rotations are ignored, and the focus is on the local rate of displacement and rotation.
 
 Buckley \& Schlegel's embedding equation is
 \begin{align}
   V(r, \theta, \phi) = R(r) A^T(r) U(\theta, \phi) + \Delta(r) ~,
   \showlabel{EmbedEq}
 \end{align}
 where $V$ is a point in $\E^4$ and $U$ is a unit sphere,
 \begin{align}
   V = \begin{pmatrix} X \\ Y \\ Z \\ W \end{pmatrix} ~,~~~~~~
   U = \begin{pmatrix} \sin\theta \cos\phi \\ \sin\theta \sin\phi \\ \cos\theta \\ 0 \end{pmatrix} ~,
 \end{align}
 while the rotation matrix $A$, and the displacement vector $\Delta$, are functions of $r$ that satisfy the following differential equations (DEs)
 \begin{align}
   A'(r) & = \Omega'(r) A(r) = 
      \begin{pmatrix}
      0 & 0 & \dfrac{P'}{S} & \dfrac{P'}{S} \alpha \\[2mm]
      0 & 0 & \dfrac{Q'}{S} & \dfrac{Q'}{S} \alpha \\[2mm]
      - \dfrac{P'}{S} & - \dfrac{Q'}{S} & 0 & \dfrac{S'}{S} \alpha \\[2mm]
      - \dfrac{P'}{S} \alpha & - \dfrac{Q'}{S} \alpha & - \dfrac{S'}{S} \alpha & 0
      \end{pmatrix}
      A(r) ~,
      \showlabel{ArDE}
      \\
   \Delta'(r) & = A^T(r)D'(r) = 
      A^T(r) \begin{pmatrix} R \dfrac{P'}{S} \\[2mm] R \dfrac{Q'}{S} \\[2mm] R \dfrac{S'}{S} \\[2mm] R' \alpha \end{pmatrix} ~,
      \showlabel{DeltarDE}
 \end{align}
 where $\alpha$ is given by \er{alphaDef}, and here too one may specify that it have the same sign as $R'$.  
 The intrinsic 3-metric of the embedded 3-surface is derived from the differential of $V$, using \er{EmbedEq}-\er{DeltarDE},
 \begin{align}
   \d s^2 & = g(\d V, \d V) = \d V^T \, \d V \\
   \d V & = (R' \, \d r) A^T U + R ((A^T)' \, \d r) U + R A^T (U_\theta \, \d\theta + U_\phi \, \d\phi) + \Delta' \, \d r \\
   & = R' A^T U \, \d r + R (A^T (\Omega')^T) U \, \d r + R A^T (U_\theta \, \d\theta + U_\phi \, \d\phi) + A^T D' \, \d r \\
   & = A^T \big[ (R' U + R (\Omega')^T U + D') \d r + R U_\theta \, \d\theta + R U_\phi \, \d\phi \big]
      \showlabel{dV} \\
   \d s^2 & = \big[ (R' U^T + R U^T \Omega' + (D')^T) \d r + R U_\theta^T \, \d\theta + R U_\phi^T \, \d\phi \big] \nn \\
      &~~~~~~ \big[ (R' U + R (\Omega')^T U + D') \d r + R U_\theta \, \d\theta + R U_\phi \, \d\phi \big] \\
   & = \d r^2 \big\{ R'^2 U^T U + R R' \big[ U^T (\Omega')^T U + U^T \Omega' U \big] + R' \big[ U^T D' + (D')^T U \big] \nn \\
      &~~~~~~~~~ + R^2 U^T \Omega' (\Omega')^T U + R \big[ U^T \Omega' D' + (D')^T (\Omega')^T U \big] + (D')^T D' \big\} \nn \\
      &~~~ + \d r \d\theta \big\{ R R' \big[ U^T U_\theta + U_\theta^T U \big]
         + R^2 \big[ U^T \Omega' U_\theta + U_\theta^T (\Omega')^T U \big] \nn \\
      &~~~~~~~~~ + R \big[ (D')^T U_\theta + U_\theta^T D' \big] \big\} \nn \\
      &~~~ + \d r \d\phi \big\{ R R' \big[ U^T U_\phi + U_\phi^T U \big]
         + R^2 \big[ U^T \Omega' U_\phi + U_\phi^T (\Omega')^T U \big] \nn \\
      &~~~~~~~~~ + R \big[ (D')^T U_\phi + U_\phi^T D' \big] \big\} \nn \\
      &~~~ + \d\theta^2 \big\{ R^2 U_\theta^T U_\theta \big\} 
         + \d\theta \d\phi \big\{ R^2 \big[ U_\theta^T U_\phi + U_\phi^T U_\theta \big] \Big\}
         + \d\phi^2 \big\{ R^2 U_\phi^T U_\phi \big\}
 \end{align}
 where
 \begin{align}
   & U_\theta = \begin{pmatrix} \cos\theta \cos\phi \\ \cos\theta \sin\phi \\ - \sin\theta \\ 0 \end{pmatrix} ~,~~~~~~~~
      U_\phi = \begin{pmatrix} - \sin\theta \sin\phi \\ \sin\theta \cos\phi \\ 0 \\ 0 \end{pmatrix}
 \end{align}
 Evaluating the various matrix products, we find
 \begin{subequations}
 \begin{align}
   & 0 = U^T \Omega' U = U^T (\Omega')^T U = U^T U_\theta = U_\theta^T U = U^T U_\phi = U_\phi^T U
      = U_\theta^T U_\phi = U_\phi^T U_\theta \\
   & 1 = U^T U = U_\theta^T U_\theta \\
   & U_\phi^T U_\phi = \sin^2\theta \\
   & (D')^T U = U^T D' = \frac{R}{S} \big\{ \sin\theta \big( P' \cos\phi + Q' \sin\phi \big) + S' \cos\theta \big\} \\
   & (D')^T (\Omega')^T U = U^T \Omega' D' =
         \frac{R}{S^2} \big\{ S' \sin\theta \big( P' \cos\phi + Q' \sin\phi \big)
         - \big( P'^2 + Q'^2 \big) \cos\theta \big\} \nn \\
      &~~~~~~~~ + \frac{R' \alpha^2}{S} \big\{ \sin\theta \big( P' \cos\phi + Q' \sin\phi \big) + \cos\theta S' \big\} \\
   & (D')^T D' = \frac{R^2}{S^2} \big\{ P'^2 + Q'^2 + S'^2 \big\} + R'^2 \alpha^2 \\
   & U^T \Omega' (\Omega')^T U = \frac{\alpha^2}{S^2} \Big(
         \big\{ S' \cos\theta + \sin\theta \big( P' \cos\phi + Q'\sin\phi \big) \big\}^2 \nn \\
      &~~~~~~~~ + \frac{\sin^2\theta}{S^2} \big\{ P' \cos\phi + Q' \sin\phi \big\}^2
         + \cos^2\theta \big\{ P'^2 + Q'^2 \big\} \Big) \\
   & U^T \Omega' U_\theta = U_\theta^T (\Omega')^T U = - \frac{\big\{ P' \cos\phi + Q' \sin\phi \big\}}{S} \\
   & U^T \Omega' U_\phi = U_\phi^T (\Omega')^T U = \frac{\cos\theta \sin\theta \big\{ P' \sin\phi - Q' \cos\phi \big\}}{S} \\
   & (D')^T U_\theta = U_\theta^T D' = \frac{R}{S} \big\{ \cos\theta \big( P' \cos\phi + Q' \sin\phi \big)
         - S' \sin\theta \big\} \\
   & (D')^T U_\phi = U_\phi^T D' = - \frac{R \sin\theta}{S} \big\{ P' \sin\phi - Q' \cos\phi \big\}
 \end{align}
 \end{subequations}
and we recover the metric \er{ds2SzAng} of the angular form of the quasi-spherical Szekeres metric.

 \subsection{Visualising the Local Embedding Geometry}
 \showlabel{VisEmb}

 Although there is no physical significance to the embedding of a given spacetime into one of higher dimension, it can be very helpful for visualising the spacetime geometry, and that is what we explore here.

 We note that the constant $w$ projection of $D'$, \er{DeltarDE}, is anti-parallel to the  $(x,y,z)$ direction of the dipole maximum, \er{xyzm}.  
 For the local rate-of-rotation matrix $\Omega'$, the fixed point locus is a 2-plane,
 \begin{align}
   \Omega' \: V = 0 ~~~~~~\to~~~~~~
   V_{fp} =
   \begin{pmatrix}
   S' \lambda \\ S' \mu \\[1mm] - (P' \lambda + Q' \mu) \\[2mm] \dfrac{(P' \lambda + Q' \mu)}{\alpha}
   \end{pmatrix}
      ~,~~~~ \lambda, \mu ~\mbox{independent parameters},
 \end{align}
 which makes $\Omega'$ the derivative of a simple rotation (not a double or isoclinic rotation).  
 The constant $w$ projections of $D'$ and $V_{fp}$ are orthogonal.  This last fact ensures the maximum tilt-displacements occur along the dipole axis (where the max \& min are located) --- see fig \ref{SphDspTlt} --- and thus enables the following argument.

 In the local $(x,y,z,w)$ frame, the constant $r$ 2-sphere lies in the $x$-$y$-$z$ 3-space, and the direction $o = (0, 0, 0, 1)$ is orthogonal to that space.  In going from the 2-sphere at $r$ to the 2-sphere at $r + \delta r$, the displacement of the sphere centres is $D' \, \delta r$; the rate of perpendicular displacement is $R' \alpha$, and the rates of sideways displacement are $R P'/S$, $R Q'/S$, \& $R S'/S$, towards the $+$ve $x$, $y$, \& $z$ directions%
 \footnote{\sf We here assume that $P'$, $Q'$, \& $S'$ are positive.  Where the opposite sign occurs, the relevant direction is changed in the obvious way.}%
 .  The slant angle, between $o$ and the line of centres, is
 \begin{align}
   \cos\gamma = o \cdot \frac{D'}{|D'|} = \frac{R'\alpha}{\sqrt{R^2 \cD_m^2 + R'^2 \alpha^2}\;}
 \end{align}
 which we could re-write as `component' angles%
 \footnote{\sf
 These `components' are the projections of $\gamma$ onto the $x$-$w$, $y$-$w$, and $z$-$w$ planes, such that $\tan^2\gamma = \tan^2\gamma_x + \tan^2\gamma_y + \tan^2\gamma_z$.
 }
 \begin{align}
   \tan\gamma_x = \frac{R P'}{R'\alpha S} ~,~~~~~~ 
   \tan\gamma_y = \frac{R Q'}{R'\alpha S} ~,~~~~~~ 
   \tan\gamma_z = \frac{R S'}{R'\alpha S} ~.
   \showlabel{gammaComps}
 \end{align}
 Since the total rotation used in \er{EmbedEq} is $A^T$, we see from \er{ArDE} that the local rotation-rate matrix to examine is $(\Omega')^T$.  
 The components of the local rate-of-shell-rotation are thus
 \begin{align}
   \omega_{xz}' \, \delta r = \frac{- P'}{S} \, \delta r ~,~~~~~~
   \omega_{yz}' \, \delta r = \frac{- Q'}{S} \, \delta r ~,~~~~~~
   \omega_{xy}' \, \delta r = 0 ~,
 \end{align}
 with senses as noted above, and the components of rate-of-tilt are
 \begin{align}
   \zeta_{xw}' \, \delta r = \frac{- P' \alpha}{S} \, \delta r ~,~~~~~~
   \zeta_{yw}' \, \delta r = \frac{- Q' \alpha}{S} \, \delta r ~,~~~~~~
   \zeta_{zw}' \, \delta r = \frac{- S' \alpha}{S} \, \delta r ~.
   \showlabel{zetaComps}
 \end{align}
 Therefore the rates of tilt are $\alpha/R$ times the rates of sideways displacement, and a 3-plane parallel to $w = 0$ is tilted down on the $-x$, $-y$, \& $-z$ sides by $\zeta_{xw}$, $\zeta_{yw}$, \& $\zeta_{zw}$, respectively.

 Both these effects, the centre displacement and the tilt, decrease the shell separation where $\cD$ has the same sign as $R'$ (the 'closing edge') and increase it where $\cD$ has the opposite sign (the 'opening edge').  The no-shell-crossing conditions ensure the separation stays positive all round each sphere.  Fig \ref{SphDspTlt} illustrates the arrangement for the case when $R' > 0$.  Although finite displacements are shown, one should think of $\delta r$ as infinitesimal, so that only first order effects are relevant.  The four displacements shown are:
 \begin{subequations}
 \begin{align}
   \mbox{the $w$-separation of shell centres} & = \delta w_0 = R' \alpha \, \delta r ~,   \showlabel{deltaw0} \\
   \mbox{the increase in shell radius} & = \delta d_0 = R' \alpha \tan\psi \, \delta r = R' \, \delta r ~, \\
   \mbox{the tilt down displacement where $\cD$ is max} & = \delta w_1 = R \zeta' \, \delta r = R \alpha \cD_m \, \delta r ~, \\
   \mbox{the dipole displacement of the shell centre} & = \delta d_1 = R \cD_m \, \delta r ~.   \showlabel{deltad1}
 \end{align}
 \showlabel{Deltas}%
 \end{subequations}
Interestingly, $\delta d_0$ \& $\delta w_0$ make the same angle as $\delta d_1$ \& $\delta w_1$.  This angle coincidence is possibly the reason why the calculation in appendix C of PG actually works --- because the displaced \& tilted shell ar $r + \delta r$ more or less lies on the LT tangent cone to the 3-surface at $r$, to first order.

 \centerline{
 \pb{145mm}{
 \centerline{
 \psset{unit=1mm, xunit=1mm, yunit=1mm}
 \pspicture*(1,16)(145,54)     
 \psset{linewidth=0.8pt,linecolor=green}
 \psline(120,20)(136,52)
 \psline(20,20)(10,40)
 \pscircle*(70,20){0.5}
 \psellipse(70,20)(50,2.5)
 \pscircle*(70,40){0.5}
 \psellipse(70,40)(60,3)
 \rput[rt](19,19){max}                 
 \rput[lt](121,19){min}
 \psset{linecolor=cyan}
 \pscircle*(76,40){0.5}
 \rput{11.2}(76,40){\psellipse(0,0)(61.2,3.051)}
 \psline[linecolor=magenta,linewidth=0.55pt,linestyle=dashed,dash=0.5 0.5](75.5,42.5)(76,40)(76.5,37.5)
 \psset{linewidth=1pt,linecolor=blue}
 \psline{->}(70,20)(70,40)                      
 \rput[r](68.5,28){$\delta w_0$}
 \psset{linecolor=red}
 \psline{->}(70,40)(76,40)
 \rput[b](72,41.5){$\delta d_1$}
 \psline[linestyle=dashed,linecolor=cyan](70,20)(76,40)
 \rput(71.5,31){$\gamma$}
 \psset{linecolor=blue}
 \psline{->}(20,20)(20,40)                      
 \rput[l](21.5,35){$\delta w_0$}
 \psline{->}(20,40)(10,40)
 \rput[b](15,41.5){$\delta d_0$}
 \psset{linecolor=red}
 \psline{->}(10,40)(10,28)
 \rput[r](8.5,34){$\delta w_1$}
 \psline{->}(10,28)(16,28)
 \rput[t](13,26.5){$\delta d_1$}
 \psset{linecolor=blue}
 \psline{->}(120,20)(120,40)                      
 \rput[r](118.5,30){$\delta w_0$}
 \psline{->}(120,40)(130,40)
 \rput[b](125,41.5){$\delta d_0$}
 \psset{linecolor=red}
 \psline{->}(130,40)(136,40)
 \rput[t](133,38.5){$\delta d_1$}
 \psline{->}(136,40)(136,52)
 \rput[l](137.5,46){$\delta w_1$}
 \rput(122,29){$\psi$}
 \endpspicture
 }
 \refstepcounter{figure}
 {\small Fig \arabic{figure}\showlabel{SphDspTlt}.~~ Sketch of the displacement and tilt effects in the embedding, with one dimension suppressed.  Because the sketch shows finite displacements, instead of infinitesimal ones, the distances shown are not exact.  The green shows the embedding of the underlying LT model --- the shells at $r$ (lower) and $r + \delta r$ (upper) and the local tangent cone.  The blue vectors show the $w$ displacement between the two shells and the radial expansion of the second shell (assumed positive here).  The cyan shows the displaced and tilted shell at $r + \delta r$ of the Szekeres model.  The red vectors show the sideways displacement of the shell centre, and the down or up displacement due to the tilt.  The dipole maximum is located at the left side, and the minimum at the right.  The $E' = 0$ locus, the `widest' part of the 2-sphere at $r + \delta r$ is shifted by $\delta d_1$; this is consistent --- a tilted slice through the LT cone has a displaced `greatest width'.  
 The tilt axis is the magenta dashed line, and it is perpendicular to the red displacement $\delta d_1$.  The $\theta$-$\phi$ rotation is not shown.
 }
 }
 }

 Looking at the next shell pair, at $r + \delta r$ and $r + 2 \delta r$, the local LT tangent cone is now tilted by $\zeta$, and the $r + 2 \delta r$ shell is tilted further.  The slant angle $\gamma$ is also tilted by $\zeta$.

   A reflection in the bottom plane gives an idea of the $R' < 0$ case.  Where $R' < 0$, $\alpha$ also flips sign and $\cD_m$ goes to $-\cD_m$ in \er{Deltas}.

 \subsection{Origins, Extrema \& Self-Intersections}
 \showlabel{OrExSfIs}

 Using the expressions in \er{Deltas}, we examine the limiting values of these embedding quantities near origins and spatial extrema.  We assume there are no shell crossings.  If there are, then $R'$ \& $\alpha$ do not always flip signs together.  We also assume a well behaved $r$ coordinate, with $R$ finite \& non-zero everywhere except at an origin, $r = r_o$, and $R'$ finite \& non-zero everywhere except at a spatial extremum, $r = r_e$.  We further assume `generic' arbitrary functions, so that, for example, $R'$ \& $\cD$ go linearly through zero at an extremum%
\footnote{\sf  
One may intentionally choose functions that give other behaviour at specific locations, such as $R'$ or $\cD$ going quadratically to zero and not changing sign.
}%
 .  The results are gathered in table \ref{FlipTbl}.
 \\[5mm]
 \noindent
 \begin{tabular}{l|c|c|c|c}
   & flips sign with     & flips sign  & behaviour at             & behaviour     \\
   & $\cD_m \to -\cD_m$  & with $R'$   & $R' \to 0 \la 1/\alpha$  & at $R \to 0$  \\ \hline
 $R$      & & & $R ~\to~ $const                         & $\to~0$ as $(r-r_o)$ \\
 $R'$     & & & $R' ~\to~ 0$ as $(r-r_e)$               & $\to~$const $\ne 0$  \\
 $\alpha$ & & & $\alpha ~\to~ \infty$ as $(r-r_e)^{-1}$ & $\to~0$ as $(r-r_o)$ \\
 $\cD_m$  & & & $\cD_m ~\to~ 0$ as $(r-r_e)$            & $\to~$const (or $0$) \\ \hline
 $\delta w_0$ & No  & No  & $R' \alpha ~\to~$const         & $\to~0$ as $(r-r_o)$    \\
 $\delta d_0$ & No  & Yes & $R' ~\to~ 0$ as $(r-r_e)$      & $\to~$const $\ne 0$     \\
 $\delta w_1$ & Yes & Yes & $R \alpha \cD_m ~\to~$const    & $\to~ 0$ as $(r-r_o)^2$ \\
 $\delta d_1$ & Yes & No  & $R \cD_m ~\to~ 0$ as $(r-r_e)$ & $\to~ 0$ as $(r-r_o)$   \\ \hline
 dipole slant $\delta d_1/\delta w_0$ & Yes & No  & $(R \cD_m)/(R' \alpha) ~\to~ 0$ as $(r-r_e)$ & $\to~$const           \\
 tilt rate $\delta w_1/R$             & Yes & Yes & $\alpha \cD_m ~\to~$const                    & $\to~ 0$ as $(r-r_o)$ \\ \hline       
 \end{tabular}
 \\[1mm]
 \refstepcounter{table}
 {\small Table \arabic{table}\showlabel{FlipTbl}.~~ The behaviour of embedding displacements near origins and spatial extrema.  See \er{Deltas} and the illustration in fig. \ref{SphDspTlt}.
 }

 Now, at an origin, where $R(t,r_o) \to 0 ~\forall~ t$, we see that the dipole slant does not disappear, but the tilt-rate must disappear.

 In contrast, at a spatial extremum, where $R'(t,r_e) \to 0 ~\forall~ t$, the dipole slant must disappear, whereas the tilt rate does not disappear.

 The condition for no local self-intersection (of adjacent shells), referring to fig \ref{SphDspTlt} and \er{Deltas}, is just:
 \begin{align}
   | \delta w_1 | & < | \delta w_0 | ~, \nn \\
   | R \alpha \cD_m \, \delta r | & < | R' \alpha \, \delta r | ~, \nn \\
   \cD_m & < \left| \frac{R'}{R} \right| ~.
 \end{align}
 Interestingly, this is just the no-shell-crossing conditions for $S$, $P$ \& $Q$ \cite{HelKra02,Hell09}.

 \subsection{Centre-Line Curvature}
 \showlabel{CnLnCv}

We now consider the locus of 2-sphere centres in the embedding space; that is, in the $(X, Y, Z, W)$ frame.  (The previous section mostly used the local $(x, y, z, w)$ frame.)

For simplicity, we will consider a model in which $P' = Q' = 0$ everywhere. Between $r$ and $r + \delta r$, the slant angle between $o$ and the line of centres changes by $\gamma_z' \, \delta r$. Part of this change is due to the fact that in the same span, the 2-spheres of constant $r$ undergo a tilt in the same direction by an angle $\zeta_{zw}' \, \delta r$, and $o$'s angle changes along with them by the same amount. The rest must be due to the change in the angle of the line of centres itself. See fig \ref{CrvCnt}. If we denote this change of the line-of-centres angle by $\delta \xi$, we can write
\begin{align}
\delta \xi &= \gamma_z' \, \delta r + \zeta_{zw}' \, \delta r
\end{align}
 \pb{30mm}{${}$\\[-5mm]
 \includegraphics[scale=0.35,natwidth=360,natheight=409]{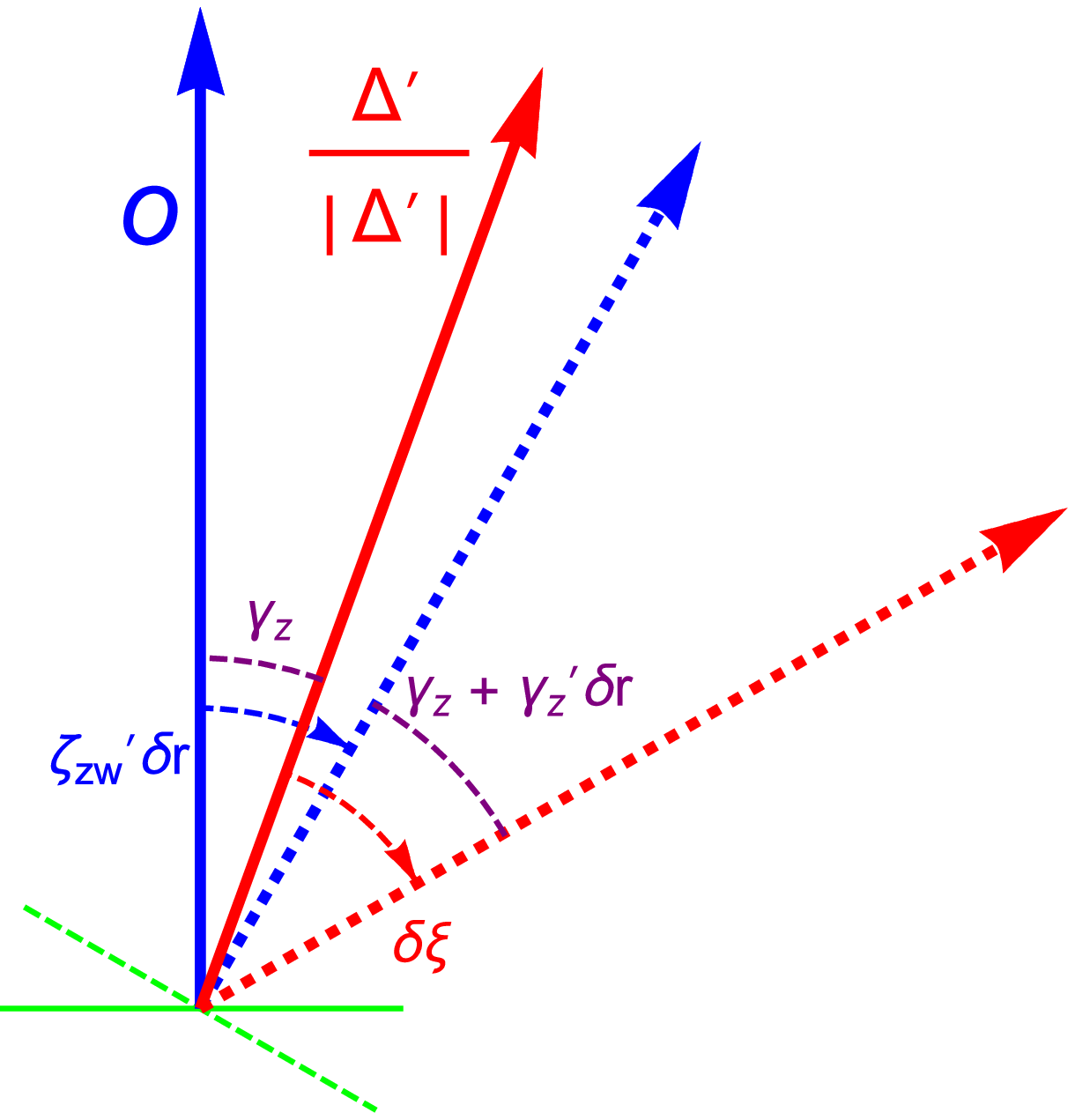}
 }
 \hfill
 \pb{120mm}{
 \refstepcounter{figure}
 {\small Fig \arabic{figure}\showlabel{CrvCnt}.~~ A diagram of the contributions to the centre line curvature.  Solid lines represent values at $r$, and dashed lines represent values at $r + \delta r$.  The green lines show the shells' orientations.  The blue lines are the vectors $o$ orthogonal to the shells, and the change between them is the rate of tilt, $\zeta_{zw}'$.  The red lines are the tangent vectors of the line of centres, which changes by $\delta \xi$.  The angles between the two, written in purple, are the slant angles $\gamma_z$.  Looking at the solid blue line and the dashed red line, we can see that the total angle between them is equal to $\gamma + \delta \xi$, but also equal to $\zeta_{zw}' \delta r + \gamma_z + \gamma_z' \delta r$, giving the result $\delta \xi = \gamma_z' \delta r + \zeta_{zw}' \delta r$.
 }
 }

Eq \er{zetaComps} gives $\zeta_{zw}'$ directly. We can calculate $\gamma_z'$ from eq \er{gammaComps} as follows:
\begin{align}
\gamma_z' \sec^2 \gamma_z &= \gamma_z' (1+\tan^2 \gamma_z) = \frac{1}{\alpha} \frac{S'}{S} - \frac{R \, R''}{R'^2 \alpha}\frac{S'}{S} - \frac{R \, \alpha'}{R' \alpha^2}\frac{S'}{S} + \frac{R}{R' \alpha}\left(\frac{S'}{S}\right)' \\
\gamma_z' &= \frac{R'^2 \alpha \frac{S'}{S} - R \, R'' \alpha \frac{S'}{S} - R \, R' \alpha' \frac{S'}{S} + R \, R' \alpha \left(\frac{S'}{S}\right)'}{R^2 \left(\frac{S'}{S}\right)^2 + R'^2 \alpha^2} ~,
\end{align}
and this can be re-expressed in terms of $\zeta_{zw}'$,
\begin{align}
\label{eq:gammaprimezeta}
\gamma_z' \sec^2 \gamma_z &= \gamma_z' (1+\tan^2 \gamma_z) = -\frac{\zeta_{zw}'}{\alpha^2} + \frac{R \, R''}{R'^2 \alpha^2}\zeta_{zw}' + 2 \frac{R \, \alpha'}{R' \alpha^3}\zeta_{zw}' - \frac{R}{R' \alpha^2}\zeta_{zw}'' \\
\gamma_z' &= \frac{-R'^2 \alpha^2 \zeta_{zw}' + R \, R'' \alpha^2 \zeta_{zw}' + 2 R \, R' \alpha \, \alpha' \zeta_{zw}' - R \, R' \alpha^2 \zeta_{zw}''}{R^2 \zeta_{zw}'^2 + R'^2 \alpha^4} ~.
\end{align}
Locally about any point, the line of centres can be approximated as following a section of a circle. The radius of this circle is determined by dividing the length traversed in a small span by the corresponding change in angle $\delta \xi$. We therefore need to express $\delta \xi$ in terms of path length along the line, rather than in terms of $\delta r$. The distance $\delta l$ between the centre of shell $r$ and that of shell $r + \delta r$ follows from the displacement components given in eqs \er{deltaw0} and \er{deltad1}:
\begin{align}
\delta l &= \delta r \sqrt{R^2 \left(\frac{S'}{S}\right)^2 + R'^2 \alpha ^2}
\end{align}
The local radius of curvature of the line of centres is then
\begin{align}
\rho_{zw} &= \frac{\delta l}{\delta \xi} \nonumber \\
&= \frac{\left[ R^2 \left(\frac{S'}{S}\right)^2 + R'^2 \alpha^2 \right]^{3/2}}{R'^2 (\alpha - \alpha^3) \frac{S'}{S} - R \, R'' \alpha \frac{S'}{S} - R \, R' \alpha' \frac{S'}{S} + R \, R' \alpha \left(\frac{S'}{S}\right)' - R^2 \alpha \left(\frac{S'}{S} \right)^3}
\showlabel{RBrhozw}
\end{align}


The relationship between the two contributions, the rate of tilt $\gamma_z'$ and the rate of change of slant $\zeta_{zw}'$, is somewhat complicated, but becomes much simpler near an extremum in $R$.  Starting from equation (\ref{eq:gammaprimezeta}), replace $R'$ by $R_2(r - r_e)$ (where $R_2$ is the value of $R''$ at $r_e$), $\alpha$ by $\frac{\alpha_e}{r - r_e}$, and $\alpha'$ by $-\frac{\alpha_e}{(r - r_e)^2}$. This gives
\begin{align}
\gamma_z' \sec^2 \gamma_z &= -\frac{\zeta_{zw}'}{\alpha_e^2}(r-r_e)^2 - \frac{R}{R_2 \alpha_e^2} \zeta_{zw}' - \frac{R}{R_2 \alpha_e^2}(r - r_e) \zeta_{zw}''
\end{align}
Since by table \ref{FlipTbl} $\zeta_{zw}'$ approaches a constant at the spatial extremum, $\zeta_{zw}''$ is small there, so we can drop the last term, as well as the first. The middle term dominates, and its sign depends only on $R_2$ and $\zeta_{zw}'$.  Therefore, $\gamma_z'$ and $\zeta_{zw}'$ have the same sign near maxima, where $R''$ is negative, and opposite signs near minima, where $R''$ is positive.  Consequently, the slant and tilt effects both increase the curvature of the line of centres near a spatial maximum, but oppose each other near a spatial minimum%
\footnote{
\sf This was first observed in numerical output, which also indicates the tilt-rate is the larger contribution in determining the centre-line curvature.}%
.

 \section{Comparison of Rotations in PG \& FR}

 In order to connect the shell rotations found in PG with the frame rotation effect found in FR, let us now set up the orthonormal basis of FR, within the $\E^4$ of section \ref{BEmbed}.  Since the basis orientation is time-independent, we will only look at spatial components.  Thus there will be 3 vectors intrinsic to the embedded surface, and one perpendicular to it.  We shall need to refer to 3 different fames; the Cartesian coordinates of $\E^4$, the Szekeres angular coordinates, and the orthonormal tetrad of the embedded surface.  Our index convention is 
 \begin{align}
   i,j,k,\cdots ~~&-~~\mbox{Szekeres coordinates}~ r,\theta,\phi \nn \\
   s,u,v,\cdots ~~&-~~\mbox{4-d flat coordinates}~ X,Y,Z,W \\
   (a),(b),(c),\cdots ~~&-~~\mbox{ortho-normal basis indices}~ (\theta),(\phi),(n),(N) ~. \nn
 \end{align}

 \subsection{Orthonormal Basis}

 From \er{dV}, the mapping between $(r, \theta, \phi)$ and $V = (X, Y, Z, W)$ consists of
 \begin{align}
   \pd{V}{r} & = A^T (R' U + R (\Omega')^T U + D') ~,~~~~
      \pd{V}{\theta} = A^T R U_\theta ~,~~~~
      \pd{V}{\phi} = A^T R U_\phi ~, \\
   \to~~~~~~~~ \Lambda^s_i & = \pd{(X,Y,Z,W)}{(r, \theta, \phi)} = 
   \begin{pmatrix}
   \ds \pd{V}{r} & \ds \pd{V}{\theta} & \ds \pd{V}{\phi}
   \end{pmatrix}
 \end{align}
 which also constitute vectors in the 3-surface along the $r$, $\theta$ \& $\phi$ directions.  Thus one may easily find a vector orthogonal to that 3-surface,
 \begin{align}
   \ol{N} & = A^T
   \begin{pmatrix}
   \alpha \sin\theta \cos\phi \\ \alpha \sin\theta \sin\phi \\ \alpha \cos\theta \\ -1 
   \end{pmatrix}
   ~,
 \end{align}
 and consequently a third surface vector orthogonal to the $\theta$ and $\phi$ directions (as well as $\ol{N}$) is
 \begin{align}
   \ol{n} & = A^T
   \begin{pmatrix}
   \sin\theta \cos\phi \\ \sin\theta \sin\phi \\ \cos\theta \\ \alpha
   \end{pmatrix}
   ~.
 \end{align}
 Normalising these, we obtain an orthonormal basis and its dual, with components in $\E^4$; the components of $\be^{(c)}$ are in the columns of \er{dbvekc}, and the basis order is $(\theta),(\phi),(n),(N)$,
 \begin{align}
   e_u{}^{(c)} & = A^T
   \begin{pmatrix}
   \cos\theta \cos\phi & -\sin\phi & \sqrt{1 + f}\; \sin\theta \cos\phi & \sqrt{-f}\; \sin\theta \cos\phi \\
   \cos\theta \sin\phi & \cos\phi & \sqrt{1 + f}\; \sin\theta \sin\phi & \sqrt{-f}\; \sin\theta \sin\phi \\
   -\sin\theta & 0 & \sqrt{1 + f}\; \cos\theta & \sqrt{-f}\; \cos\theta \\
   0 & 0 & \sqrt{-f}\; & -\sqrt{1 + f}\;
   \end{pmatrix}
   = A^T \ol{e}_u{}^{(c)} ~,
   \showlabel{dbvekc} \\
   e_{(b)}{}^s & =
   \begin{pmatrix}
   \cos\theta \cos\phi & \cos\theta \sin\phi & - \sin\theta & 0 \\
   - \sin\phi & \cos\phi & 0 & 0 \\
   \sqrt{1 + f}\; \sin\theta \cos\phi & \sqrt{1 + f}\; \sin\theta \sin\phi & \sqrt{1 + f}\; \cos\theta & \sqrt{-f}\; \\
   \sqrt{-f}\; \sin\theta \cos\phi & \sqrt{-f}\; \sin\theta \sin\phi & \sqrt{-f}\; \cos\theta & -\sqrt{1 + f}\; 
   \end{pmatrix}
   A
   = \ol{e}_{(b)}{}^s A ~.
   \showlabel{bvebj}
 \end{align}
 Equation \er{bvebj} gives the flat-space components of $\be_{(b)}$ in $\E^4$.

 \subsection{Variation of the Embedded Basis}

 Consider a path parametrised by $\lambda$, within the embedded 3-surface, which is also a 3-space of constant time in the Szekeres metric, so that $r = r(\lambda)$, $\theta = \theta(\lambda)$, $\phi = \phi(\lambda)$.  The tangent vector is $v^j = \tdil{x^j}{\lambda}$, and the path in $\E^4$ is $V = V(r(\lambda), \theta(\lambda), \phi(\lambda))$.  Along this path, the variation of the flat space basis vector $\be_{(b)}$ within $\E^4$ is 
 \begin{align}
   \nabla_{\mb v} \be_{(b)} & = v^s \, \nabla_{\be_s} \big( e_{(b)}{}^u \be_u \big)
         = v^s \big( e_{(b)}{}^u{}_{,s} \, \be_u + e_{(b)}{}^u \nabla_{\be_s} \, \be_u \big)
         = v^s \, e_{(b)}{}^u{}_{,s} \, \be_u ~,
 \end{align}
 so that the frame rotation matrix of FR is (see eq (21) of that paper)
 \begin{align}
   {\cal V}^{(c)}{}_{(b)} & = \big( \nabla_{\mb v} \, \be_{(b)} \big) (\be^{(c)})
          = \big( v^s \, e_{(b)}{}^u{}_{,s} \, \be_u \big) \big( e_v{}^{(c)} \, \be^v \big)
          = v^s \, e_{(b)}{}^u{}_{,s} \, e_u{}^{(c)} \nn \\
   & = v^i \, \Lambda^s_i \, e_{(b)}{}^u{}_{,s} \, e_u{}^{(c)}
          = v^i \, e_{(b)}{}^u{}_{,i} \, e_u{}^{(c)} ~.
 \end{align}
 We have
 \begin{align}
 \begin{aligned}
   e_{(b)}{}^u{}_{,r} & = \ol{e}_{(b)}{}^u \, A' + \ol{e}_{(b)}{}^u{}_{,r} \, A
         = \big( \ol{e}_{(b)}{}^u \, \Omega' + \ol{e}_{(b)}{}^u{}_{,r} \big) A \\
   e_{(b)}{}^u{}_{,\theta} & = \ol{e}_{(b)}{}^u{}_{,\theta} \, A ~,~~~~~~~~~~
         e_{(b)}{}^u{}_{,\phi} = \ol{e}_{(b)}{}^u{}_{,\phi} \, A ~,
 \end{aligned}
 \end{align}
 so that
 \begin{align}
   {\cal V}^{(c)}{}_{(b)} = v^i \, e_{(b)}{}^u{}_{,i} \, e_u{}^{(c)} & = \big\{
      \dot{r} \big( \ol{e}_{(b)}{}^u \, \Omega' + \ol{e}_{(b)}{}^u{}_{,r} \big)
      + \dot{\theta} \, \ol{e}_{(b)}{}^u{}_{,\theta} + \dot{\phi} \, \ol{e}_{(b)}{}^u{}_{,\phi}
     \big\} A \, A^T \, \ol{e}_u{}^{(c)} \nn \\
   & = \big\{
      \dot{r} \big( \ol{e}_{(b)}{}^u \, \Omega' + \ol{e}_{(b)}{}^u{}_{,r} \big)
      + \dot{\theta} \, \ol{e}_{(b)}{}^u{}_{,\theta} + \dot{\phi} \, \ol{e}_{(b)}{}^u{}_{,\phi}
     \big\} \ol{e}_u{}^{(c)} ~.
     \showlabel{Vcb}
 \end{align}
 The basis derivatives with respect to Szekeres angular coordinates are
 \begin{subequations}
 \begin{align}
   \ol{e}_{(b)}{}^s{}_{,r} & = \frac{f'}{2}
   \begin{pmatrix}
   0 & 0 & 0 & 0 \\
   0 & 0 & 0 & 0 \\
   \dfrac{\sin\theta \cos\phi}{\sqrt{1 + f}\;} & \dfrac{\sin\theta \sin\phi}{\sqrt{1 + f}\;}
      & \dfrac{\cos\theta}{\sqrt{1 + f}\;} & \dfrac{-1}{\sqrt{- f}\;} \\[3mm]
   \dfrac{- \sin\theta \cos\phi}{\sqrt{- f}\;} & \dfrac{- \sin\theta \sin\phi}{\sqrt{- f}\;}
      & \dfrac{- \cos\theta}{\sqrt{- f}\;} & \dfrac{-1}{\sqrt{1 + f}\;}
   \end{pmatrix}
   ~,
   \showlabel{ebsr} \\
   \ol{e}_{(b)}{}^s{}_{,\theta} & =
   \begin{pmatrix}
   - \sin\theta \cos\phi & - \sin\theta \sin\phi & - \cos\theta & 0 \\
   0 & 0 & 0 & 0 \\
   \sqrt{1 + f}\; \cos\theta \cos\phi & \sqrt{1 + f}\; \cos\theta \sin\phi & - \sqrt{1 + f}\; \sin\theta & 0 \\
   \sqrt{- f}\; \cos\theta \cos\phi & \sqrt{- f}\; \cos\theta \sin\phi & - \sqrt{- f}\; \sin\theta & 0
   \end{pmatrix}
   ~,
   \showlabel{ebsth} \\
   \ol{e}_{(b)}{}^s{}_{,\phi} & =
   \begin{pmatrix}
   - \cos\theta \sin\phi & \cos\theta \cos\phi & 0 & 0 \\
   - \cos\phi & - \sin\phi & 0 & 0 \\
   - \sqrt{1 + f}\; \sin\theta \sin\phi & \sqrt{1 + f}\; \sin\theta \cos\phi & 0 & 0 \\
   - \sqrt{- f}\; \sin\theta \sin\phi & \sqrt{- f}\; \sin\theta \cos\phi & 0 & 0
   \end{pmatrix}
   ~.
   \showlabel{ebsph}
 \end{align}
 \end{subequations}
Using eqs \er{ebsr}-\er{ebsph} and \er{dbvekc}-\er{bvebj} in \er{Vcb} above, we find that
 \begin{subequations}
 \begin{align}
   {\cal V}^{(\theta)}{}_{(n)} & =
      \dfrac{- f S' \sin\theta - [(1 + f) - f \cos\theta]\{P' \cos\phi + Q' \sin\phi\} v^r}{\sqrt{1 + f}\; S} + \sqrt{1 + f}\; v^\theta ~,
      \showlabel{Vthn} \\
   {\cal V}^{(\phi)}{}_{(n)} & = \dfrac{[1 - (1 + f)(1 - \cos\theta)]\{P' \sin\phi - Q' \cos\phi\} v^r}{\sqrt{1 + f}\; S}
      + \sqrt{1 + f}\; \sin\theta v^\phi ~,
      \showlabel{Vphn} \\
   {\cal V}^{(\phi)}{}_{(\theta)} & = \dfrac{- (P' \sin\phi - Q' \cos\phi) \sin\theta}{S} v^r + \cos\theta v^\phi ~.
      \showlabel{Vphth}
 \end{align}
 \end{subequations}
Eqs \er{Vthn}-\er{Vphth} agree with FR(25)-(27), confirming that the frame rotation effects found in \cite{Hell17} are fully consistent with, and therefore explained by, the shell rotations and tilts uncovered in \cite{BucSch13} \& \cite{BucSch19}.

 \section{Toroidal \& Rotating Embeddings}
 \showlabel{TrdMbd}

 As an illustration of the rotations and tilts described in PG \& FR, it would be interesting to see if one can define a Szekeres model whose embedding `naturally' bends round and closes on itself, i.e. the embedded surface has the topology of a torus in the 4-d flat space, without any arbitrary identifications.  It doesn't have to be a Datt-Kantowski-Sachs (DKS) type model \cite{Datt38,KanSac66}, it could be a quasi-spherical model that has both a spatial maximum and a minimum; or multiple spatial maxima \& minima.  Of course there is no physical significance to the embedding of a spacetime%
 \footnote{\sf ... unless it's a brane.}%
 , and the shape of the embedded surface depends on the space it is embedded into.  The point here is to illustrate that the tilt can be continuously in the same sense.

 Where spatial extrema occur, the conditions for no shell crossings or surface layers \cite{HelKra02,Hell09} require that $f = - 1$, all 6 arbitrary functions have zero derivative, $0 = M' = f' = a' = S' = P' = Q'$, and $R'$, $M'$, $\alpha$ must change sign together.

 In order to achieve this, we require 3 things as $r$ runs from $r_i$ to $r_f$: (i) the tilt/rotation matrix $A(r)$ must run round $2 \pi$, relative to some `axis', and return to the identity; (ii) the locus of centres $\Delta(r)$ must form a loop; (iii) the areal radius $R(r)$ and its derivative $R'(r)$ must return to their starting values, so that the join is smooth  --- for example $R$ could be (multiply) periodic in $r$ around this loop --- and this should hold at each constant $t$; (iv) ideally the 3-surface should not intersect itself in the embedding, and there should be no shell crossings.

 \subsection{Embedding DEs}

 The matrix DEs of \er{ArDE} actually separate out into 4 identical sets of 4 linked DEs, but with different initial conditions,
 \begin{subequations}
 \begin{align}
   A(r) & =
      \begin{pmatrix}
      \nu_1(r) & \nu_2(r) & \nu_3(r) & \nu_4(r) \\
      \chi_1(r) & \chi_2(r) & \chi_3(r) & \chi_4(r) \\
      \lambda_1(r)
        & \lambda_2(r)
            & \lambda_3(r)
                & \lambda_4(r) \\
      \sigma_1(r)
        & \sigma_2(r)
            & \sigma_3(r)
                & \sigma_4(r)
      \end{pmatrix}
      ~,~~~~~~ A(0) = I ~, \\
   \nu_i' & = \frac{P'}{S} (\lambda_i + \alpha \sigma_i) ~,~~~~
         \chi_i' = \frac{Q'}{S} (\lambda_i + \alpha \sigma_i) ~, \nn \\
      \lambda_i' & = \frac{- P' \nu_i - Q' \chi_i + S' \alpha \sigma_i}{S} ~,~~~~
         \sigma_i' = - \frac{\alpha (P' \nu_i + Q' \chi_i + S' \lambda_i)}{S} ~,~~~~ i = 1,2,3,4 ~; \\
   \nu_1(0) & = 1 ~,~~~ \chi_1(0) = 0 ~,~~~ \lambda_1(0) = 0 ~,~~~ \sigma_1(0) = 0 ~, \\
   \nu_2(0) & = 0 ~,~~~ \chi_2(0) = 1 ~,~~~ \lambda_2(0) = 0 ~,~~~ \sigma_2(0) = 0 ~, \\
   \nu_3(0) & = 0 ~,~~~ \chi_3(0) = 0 ~,~~~ \lambda_3(0) = 1 ~,~~~ \sigma_3(0) = 0 ~,\\
   \nu_4(0) & = 0 ~,~~~ \chi_4(0) = 0 ~,~~~ \lambda_4(0) = 0 ~,~~~ \sigma_4(0) = 1 ~.
 \end{align}
   \showlabel{A-DE-ICs}%
 \end{subequations}
Similarly, from \er{DeltarDE}, the initial value problem (IVP) for the line of shell centres separates out into the same groups,
 \begin{subequations}
 \begin{align}
   \Delta(r) & = \begin{pmatrix} X_C \\ Y_C \\ Z_C \\ W_C \end{pmatrix} ~,~~~~~~
   \Delta(0) = 0 ~, \\
   (\Delta^i)' & = \frac{R}{S} (P' \nu_i + Q' \chi_i + S' \lambda_i) + R' \alpha \sigma_i ~,~~~~ i = 1,2,3,4 ~,
 \end{align}
   \showlabel{Dl-DE-ICs}%
 \end{subequations}
making 4 identical sets of 5 DEs.

 \subsection{Case With Only $S$ Varying}

 Let's take the simple case of $P' = 0 = Q'$, while $f$, $M$, $a$, $S$ are general. \\
 (i) Then by \er{ArDE}
 \begin{align}
   A'(r) & = \Omega'(r) A(r) = 
      \begin{pmatrix}
      0 & 0 & 0 & 0 \\[2mm]
      0 & 0 & 0 & 0 \\[2mm]
      0 & 0 & 0 & \dfrac{S'}{S} \alpha \\[2mm]
      0 & 0 & - \dfrac{S'}{S} \alpha & 0
      \end{pmatrix}
      A(r)
 \end{align}
 which integrates up to
 \begin{align}
   A & =
      \begin{pmatrix}
      1 & 0 & 0 & 0 \\[2mm]
      0 & 1 & 0 & 0 \\[2mm]
      0 & 0 & \cos\zeta & \sin\zeta \\[2mm]
      0 & 0 & - \sin\zeta & \cos\zeta
      \end{pmatrix}
      ~,~~~~~~
      \zeta(r) = \int_{r_i}^{r} \frac{S' \alpha}{S} \, \d r ~.
      \showlabel{AzetaExInt}
 \end{align}
 Then from \er{AzetaExInt} and \er{alphaDef} we have
 \begin{align}
 \begin{aligned}
   f & = \frac{- S^2 \zeta'^2}{(S^2 \zeta'^2 + S'^2)}
      = \frac{- 1}{\left( 1 + \dfrac{S'^2}{S^2 \zeta'^2} \right)} ~, \\
   f & = - 1 ~~~~\to~~~~ S'^2 = 0 ~,~~~~ \zeta' \neq 0 ~, \\
   f & = - f_a ~~~~\to~~~~ S'^2 = (1 - f_a) S^2 \zeta'^2 ~.
 \end{aligned}
 \showlabel{fSzeta}
 \end{align}
 One may thus obtain all of $f(r)$, $\alpha(r)$, $S(r)$ or $\zeta(r)$ from specifying just two of them.  It should be fairly easy to make $\zeta$ run from $0$ to $2 \pi$, because it does not depend on time through $R$ or $R'$.  We show below that $f$ and hence $\alpha$ will be oscillatory, and we note that $\alpha$ diverges where $f$ goes to $- 1$, so $S' = 0$ will be needed here.  Thus $S' \alpha/S$ needs to be non-negative, which means $S'$ changes sign with $\alpha$.  Possible functional forms for $\zeta(r)$, $S(r)$, $f(r)$, and $\alpha(r)$ will be considered below.

 (ii) Next by integrating \er{DeltarDE}, and requiring that the line of centres closes up, we find
 \begin{align}
   \Delta' & = A^TD' = 
      A^T \begin{pmatrix} 0 \\[1mm] 0 \\[1mm] R \dfrac{S'}{S} \\[3mm] R' \alpha \end{pmatrix} ~, \\
   \int_{r_i}^{r_f} \Delta' \, \d r & = 0 = \int_{r_i}^{r_f} 
      \begin{pmatrix} 0 \\[1mm] 0 \\[1mm] \cos\zeta R \dfrac{S'}{S} - \sin\zeta R' \alpha \\[3mm]
                      \sin\zeta R \dfrac{S'}{S} + \cos\zeta R' \alpha \end{pmatrix}
      \d r ~.
 \end{align}
 Generically, $R$ \& $R'$ don't have the same time dependence, so, in order for this to be true at all times, we attempt to arrange that
 \begin{subequations}
 \begin{align}
   \int_{r_i}^{r_f} \cos\zeta R \dfrac{S'}{S} \, \d r = 0 = \int_{r_i}^{r_f} \sin\zeta R' \alpha \, \d r ~, \\
   \int_{r_i}^{r_f} \sin\zeta R \dfrac{S'}{S} \, \d r = 0 = \int_{r_i}^{r_f} \cos\zeta R' \alpha \, \d r ~.
 \end{align}
   \showlabel{IntZero4}%
 \end{subequations}
  In particular, if within $0 \le \zeta \le \pi$, one can arrange that $R$ goes max-to-min-to-max (or min-to-max-to-min), in a manner that's symmetric about $\zeta = \pi/2$, then, however the time evolution goes, 2 copies of this should join up nicely.  Similarly, if within $0 \le \zeta \le \pi/2$, one can arrange that $R$ goes max-to-min-to-max (or min-to-max-to-min), symmetrically about $\zeta = \pi/4$, then 4 copies of this should also join up nicely.  And so on.

 (iii) To make $R$ \& $R'$ join nicely, so that the torus `tube' joins itself smoothly, we could choose $f$, $M$ \& $a$ periodic; and of course $f = -1$ is needed at the extrema, so $f$ must oscillate twice as fast.

 (iv) Once a detailed model has been chosen, it can be checked for shell crossings and self intersections, and adjustments can be made as needed.

 \subsection{Case With Only $P$ Varying}

 Next we consider the case $S' = 0 = Q'$ (\& general $f$, $M$, $a$, $P$).  Then by \er{A-DE-ICs} \& \er{Dl-DE-ICs}, the IVP becomes
 \begin{subequations}
 \begin{align}
   \nu_i' & = \frac{P'}{S} (\lambda_i + \alpha \sigma_i) ~,~~~~
      \lambda_i' = - \frac{P'}{S} \nu_i ~,~~~~
      \sigma_i' = - \frac{P' \alpha}{S} \nu_i ~,~~~~ i = 1,2,3,4 ~; \\
   \nu_1(0) & = 1 ~,~~~ \lambda_1(0) = 0 ~,~~~ \sigma_1(0) = 0 ~, \\
   \nu_2(0) & = 0 ~,~~~ \lambda_2(0) = 0 ~,~~~ \sigma_2(0) = 0 ~, \\
   \nu_3(0) & = 0 ~,~~~ \lambda_3(0) = 1 ~,~~~ \sigma_3(0) = 0 ~,\\
   \nu_4(0) & = 0 ~,~~~ \lambda_4(0) = 0 ~,~~~ \sigma_4(0) = 1 ~, \\
   (\Delta^i)' & = \nu_i \frac{R P'}{S} + \sigma_i R' \alpha ~,~~~~~~
      \Delta(0) = 0 ~.
 \end{align}
   \showlabel{AnulamsigICs}%
 \end{subequations}
Clearly $\chi_i = 0 = \nu_2 = \lambda_2 = \sigma_2 = Y_C$ for all $r$.

 Cyclic choices for the arbitrary functions similar to those above should again ensure closure.

 \subsection{Model 1, $S$ Only}
 \showlabel{Mdl1S}

 We choose $\zeta$ to be uniformly increasing, satisfying (i) by construction, we fix $P = 0 = Q$, and we let $S$ oscillate,
 \begin{align}
   \zeta & = \mu r ~, \\
   S & = S_0 + S_1 \big\{ 1 - \cos(n \mu r) \big\} ~,~~~~~~ n \in {\mathbb N} ~,
 \end{align}
 in which case, \er{AzetaExInt}, \er{alphaDef} and \er{fSzeta} give%
 \footnote{\sf We could choose $\alpha$ to be the negative of the r.h.s. of \er{alphatorus}, which would merely reflect the embedding in the $W = 0$ plane.}
 \begin{align}
   f & = \frac{- 1}{1 + \dfrac{n^2 S_1^2 \sin^2(n \mu r)}{\big(S_0 + S_1 \big\{ 1 - \cos(n \mu r) \big\} \big)^2}} ~,
      \showlabel{ftorus} \\
   \to~~~~~~~~ \alpha & = \frac{S_0 - S_1 \big\{ 1 - \cos(n \mu r) \big\}}{n S_1 \sin(n \mu r)} ~,
      \showlabel{alphatorus}
 \end{align}
 and the maxima of $f$ are at 
 \begin{align}
   \cos(n \mu r) = \frac{S_1}{S_0 + S_1} ~~~~\to~~~~
      f_m = \frac{- 1}{1 + \dfrac{n^2 S_1^2}{S_0 (S_0 + 2 S_1)}} ~.
 \end{align}
 For the mass and bang time we choose 
 \begin{align}
   M & = M_0 + M_1 \big\{ 1 - \cos(n \mu r) \big\} ~,~~~~~~ 
   a = a_0 - a_1 \big\{ 1 - \cos(n \mu r) \big\} ~,
 \end{align}
 in accordance with requirement (iii).

 The relevant no-shell-crossing requirements are that $-a'$ and $a' + 2 \pi T'$ have the same sign as $M'$, and that $(E'/E)_m \le \big| M'/(3 M) \big|$.  The function $f$, which appears in the denominator of $T$, defined by eq \er{ScLT}, oscillates twice as fast as $M$ does.  Thus the amplitude of $f$'s variation in eq \er{ftorus} needs to be small.  For the model considered here, this means $S_1$ should be small, which in turn means $S$ does not vary much.

 Now the calculation of $R(r)$ and $R'(r)$ on a constant time slice is a numerical exercise; the $t$ equation of \er{RptpEll} has to be solved to get $\eta$ at each $r$.  But, using the parametric solution \er{RptpEll} with zero $\Lambda$, we can instead integrate $\Delta'$ on surfaces of constant $\eta$; if the result is zero for every constant-$\eta$ surface, then it will be zero for every constant-$t$ surface also.  Using constant $\eta$ surfaces has the additional advantage that the bang and crunch singularities are avoided.  The integrals in \er{IntZero4} are therefore modified to become
 \begin{subequations}
 \begin{align}
   \int_{r_i}^{r_f} \cos\zeta R \dfrac{S'}{S} \, \d r & = (1 - \cos\eta) \int_{r_i}^{r_f}
         \cos \left[ \int_{r_i}^{r} \frac{S' \alpha}{S} \, \d r  \right] \frac{M S'}{(- f) S} \, \d r ~, \\
   \int_{r_i}^{r_f} \cos\zeta R' \alpha \, \d r & = (1 - \phi_1) (1 - \cos\eta) \int_{r_i}^{r_f}
         \cos \left[ \int_{r_i}^{r} \frac{S' \alpha}{S} \, \d r  \right] \frac{M' \alpha}{(- f)} \, \d r \nn \\
      &~~~~ - \left( \frac{3}{2} \phi_1 - 1 \right) (1 - \cos\eta) \int_{r_i}^{r_f}
         \cos \left[ \int_{r_i}^{r} \frac{S' \alpha}{S} \, \d r  \right] \frac{f' M \alpha}{f^2} \, \d r \nn \\
      &~~~~ - \phi_2 (1 - \cos\eta) \int_{r_i}^{r_f}
         \cos \left[ \int_{r_i}^{r} \frac{S' \alpha}{S} \, \d r  \right] (- f)^{1/2} a' \alpha \, \d r ~,
 \end{align}
 \showlabel{CentrelineI}%
 \end{subequations}
with obvious variations for the other two.  By symmetry arguments, the 4 integrals of \er{IntZero4} clearly evaluate to zero when the above choices are made, ensuring that (ii) is satisfied.  See fig \ref{NtgPlt} for a sample plot of one of the terms in \er{CentrelineI}.

 \centerline{
 \pb{145mm}{
 \centerline{\includegraphics[angle=-90,scale=0.4]{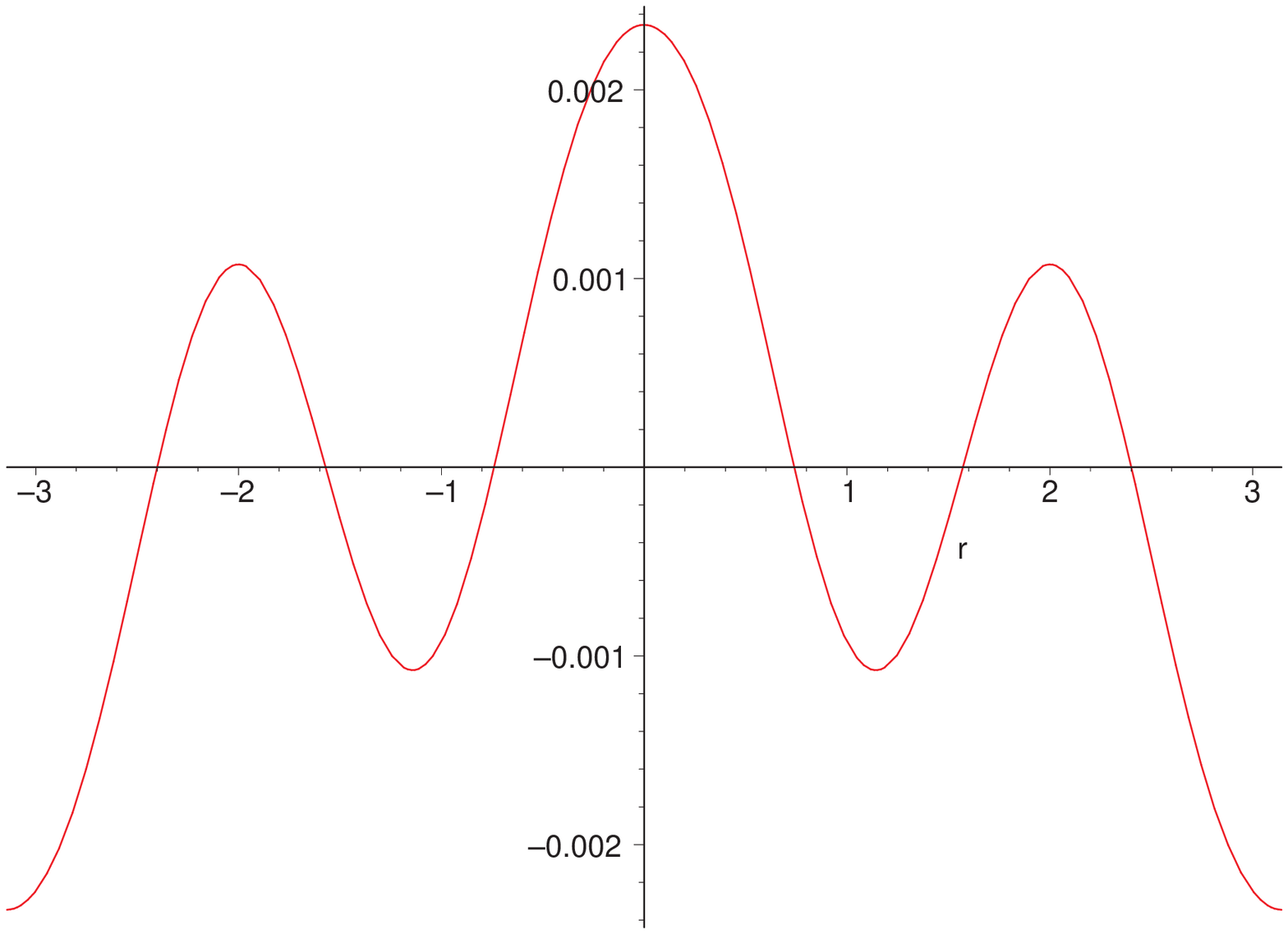}}
 \refstepcounter{figure}
 {\small Fig \arabic{figure}\showlabel{NtgPlt}.~~ The integrand $\cos\zeta (M f' \alpha/f^2)(3 \phi_1/2 - 1)(1 - \cos\eta)$ of \er{CentrelineI} at $\eta = \pi/3$, for the functions and parameter values given in section \ref{Mdl1S} and also used in fig \ref{TrsSlc}.
 }}}

 A particular example satisfying all requirements is plotted in fig \ref{TrsSlc}, for the parameter times $\eta = \pi/3$ and $\eta = 5 \pi/3$.  The parameter values are $n = 2$, $\mu = \pi$, $r_i = - 1$, $r_f = 1$, $S_0 = 1$, $S_1 = 0.1$, $M_0 = 0.1$, $M_1 = 0.05$, $a_0 = 0$, $a_1 = 0.1$.  The no-shell-crossing conditions \cite{HelLak85,HelKra02} that are applicable here are $(-a')/M' \ge0$, $(a' + 2 \pi T')/M' \ge 0$, $(E'/E)_m \le \big| M'/(3 M) \big|$; and it has been checked that the chosen functions and parameter values satisfy them, though not by much.  The construction used here puts strong limits on the variation of $S$.

 \centerline{
 \pb{145mm}{
 \centerline{
   \includegraphics[angle=-90,scale=0.4]{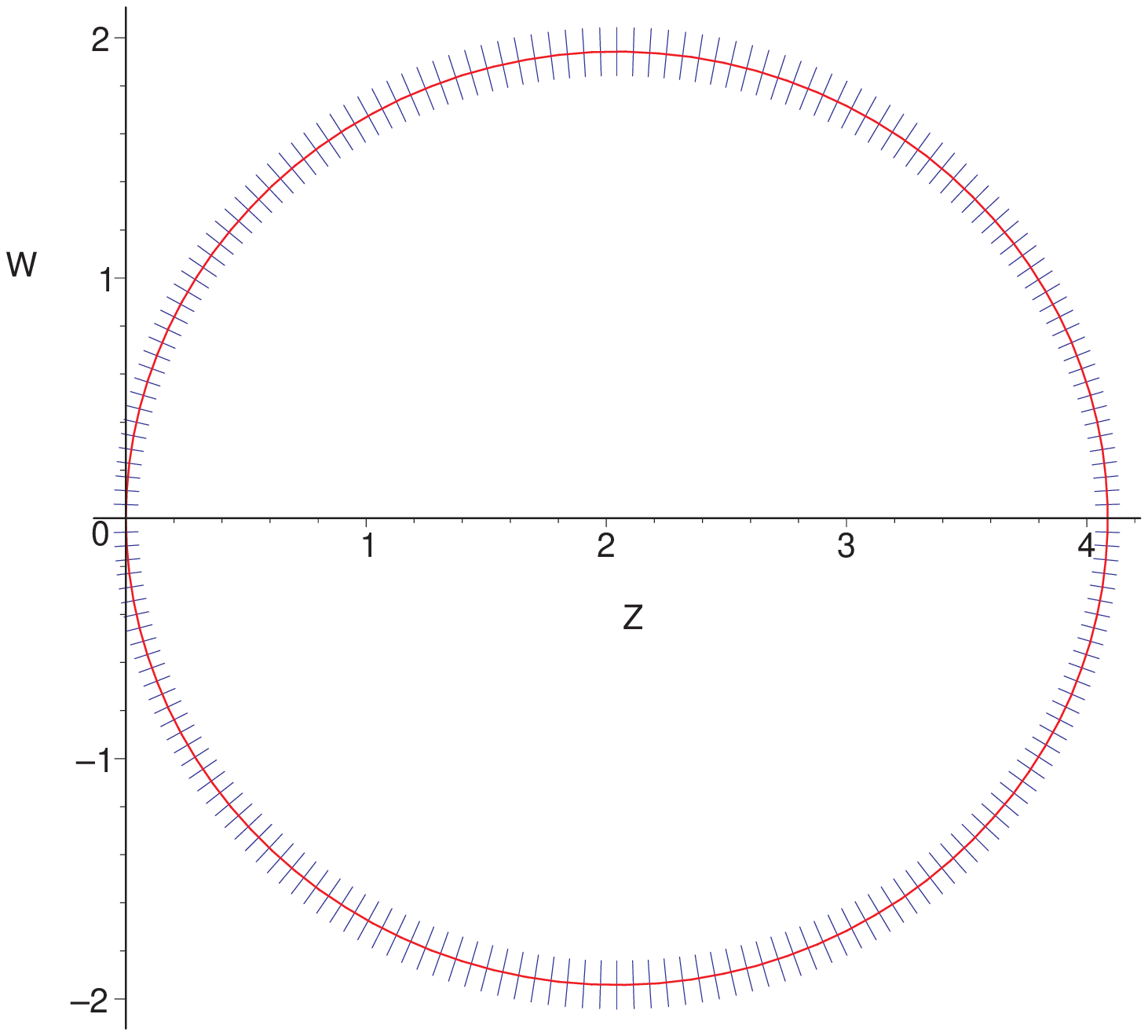}
   ~~~~
   \includegraphics[angle=-90,scale=0.4]{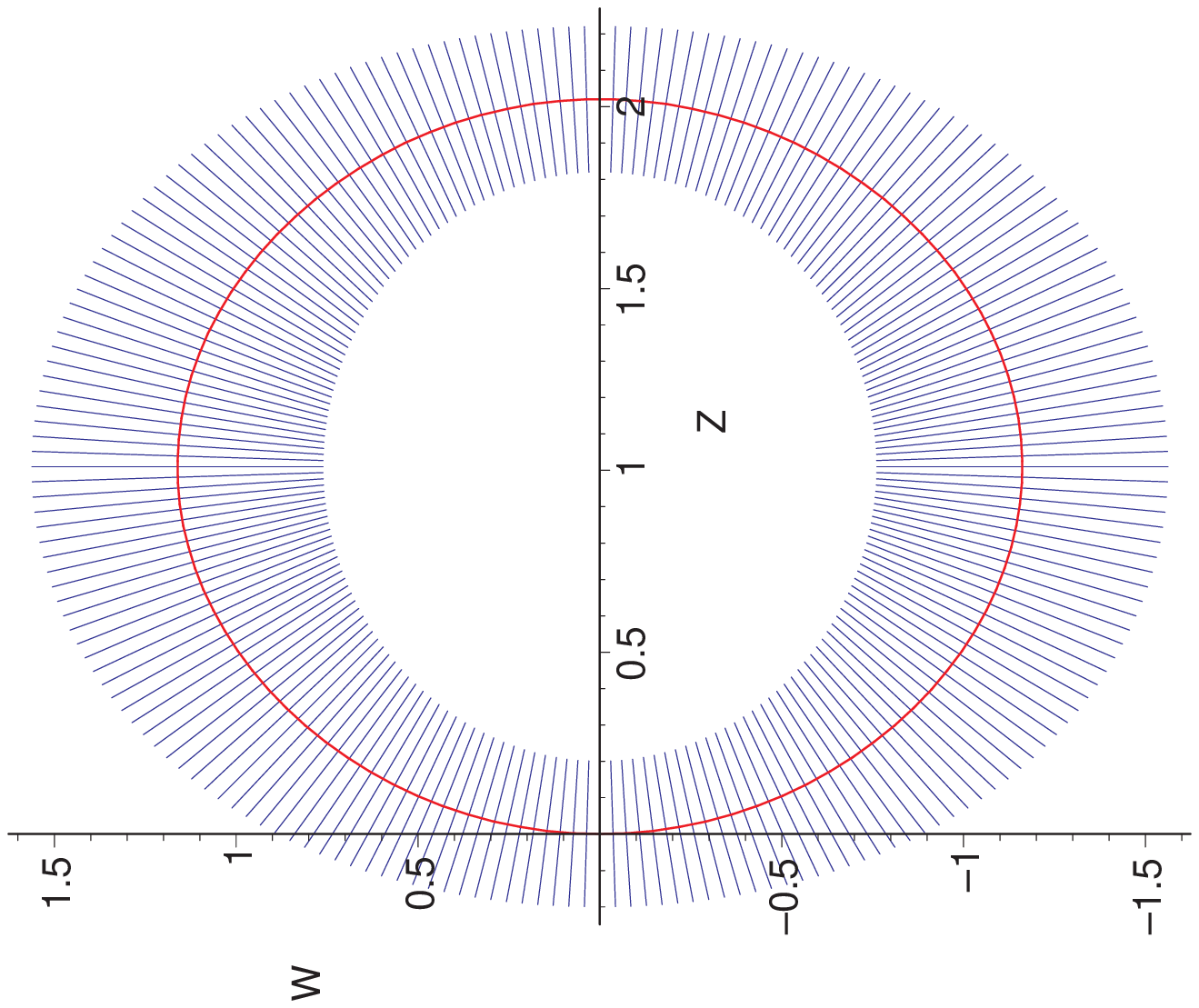}
   ~~~~
   \includegraphics[angle=-90,scale=0.4]{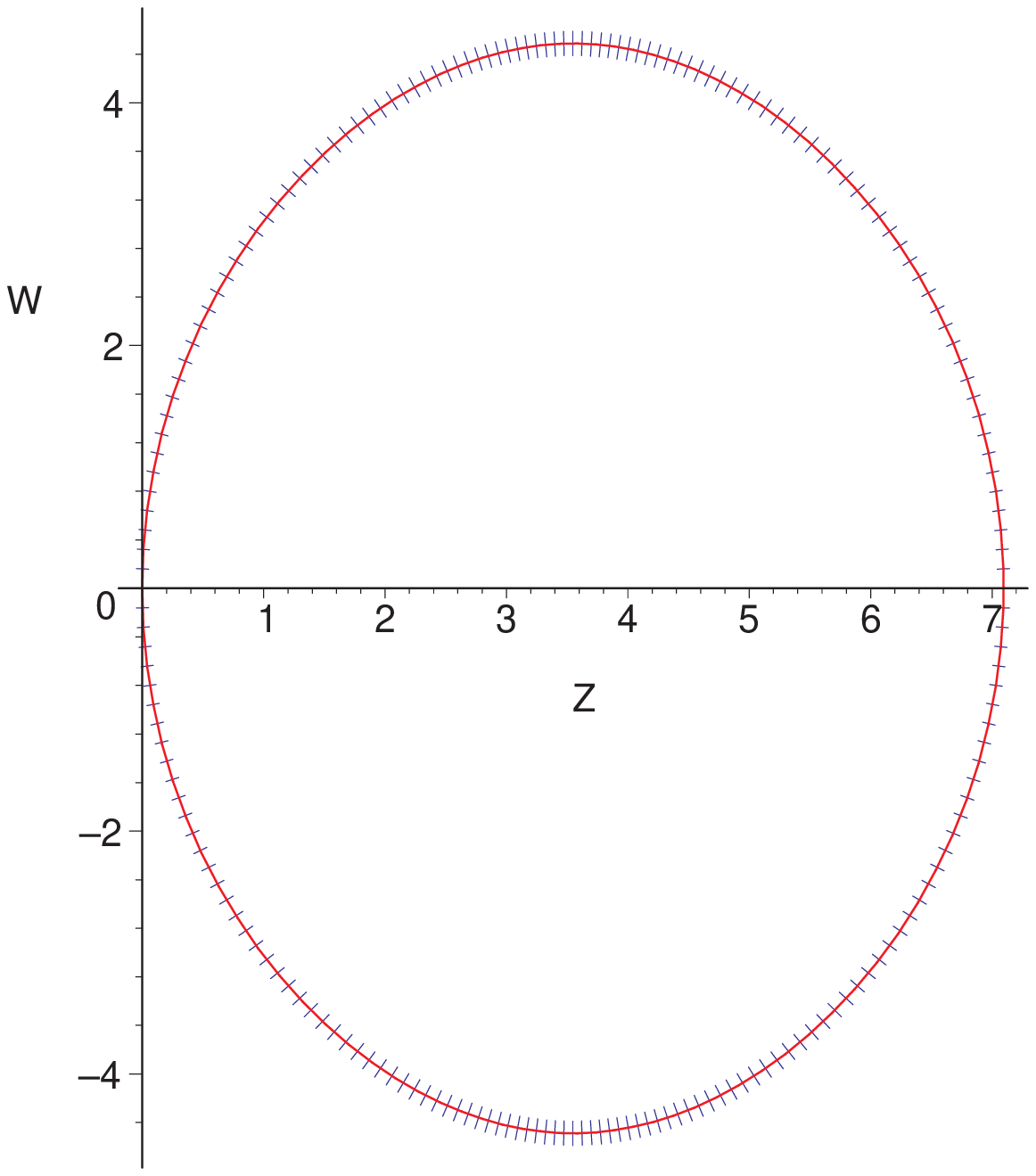}
 }
 \refstepcounter{figure}
 {\small Fig \arabic{figure}\showlabel{TrsSlc}.~~ An example of an embedded Szekeres model with a `natural' torus topology as described in section \ref{TrdMbd}.  The path of the sphere centres (red), and a selection of sphere diameters (blue), are shown in the $(Z,W)$ plane.  Each diameter represents a complete 2-sphere, and the sequence of 2-spheres describes an embedded 3-surface in the flat 4-d $(X, Y, Z, W)$ space.  The arbitrary functions and parameter values of this model are listed in section \ref{Mdl1S}.  The left plot is for $\eta = \pi/3$, the middle one for $\eta = \pi$, and the right one for $\eta = 5 \pi/3$.  These are not constant time plots, so they are indicative rather than precise.  
 }}}

 \subsection{Model 2, $S$ Only}
 \showlabel{Mdl2S}


 This is a much wobblier version of the toroidal model, that has 3 lobes instead of 2.  The key defining functions are
 \begin{align}
 \begin{aligned}
   M(r) & = M_0 + M_1 \big\{ 1 - \cos(n \mu_0 r) \big\} ~, \\
   f(r) & = - 1 + f_1 \big\{ 1 - \cos(2 n \mu_0 r) \big\} ~, \\
   a(r) & = 0 ~, \\
   \zeta' & = \mu_0 \big\{ 1 + \mu_1 \cos(n \mu_0 r) \big\} ~, \\
   S'/S & = \zeta'/\alpha ~,
 \end{aligned}
 \end{align}
 from which we find
 \begin{align}
 \begin{aligned}
   \alpha & = \frac{\sqrt{1 - 2 f_1 \sin^2(n \mu_0 r)}\;}{\sqrt{2 f_1}\; \sin(n \mu_0 r)} ~,~~~~
      \zeta = \mu_0 \left( r + \frac{\mu_1 \sin(n \mu_0 r)}{n \mu_0} \right) \\
   \frac{S'}{S} & = \frac{\sqrt{2 f_1}\; \mu_0 \sin(n \mu_0 r) \big( 1 + \mu_1 \cos(n \mu_0 r) \big)}{\sqrt{1 - 2 f_1 \sin^2(n \mu_0 r)}\;} \\
   S & = S_0 \exp \left\{ \frac{- \mu_1 \sqrt{1 - 2 f_1 \sin^2(n \mu_0 r)}\;}{\sqrt{2 f_1}\;n} \right\} \\
      &~~~~ \left[ \frac{(1 - 2 f_1 + 4 f_1 \cos^2(n \mu_0 r))}{2 \sqrt{2 f_1}\;}
         - \cos(n \mu_0 r) \sqrt{1 - 2 f_1 \sin^2(n \mu_0 r)}\; \right]^{1/(2 n)}
 \end{aligned}
 \end{align}
 We can choose the value of ~$S_0$~ to make ~$S = 1$~ at ~$r = 0$.  
 For the plots in fig \ref{Md2Plt}, the parameters are
 \begin{align}
   M_0 = 0.55 ~,~~~~ M_1 = 0.7 ~,~~~~ f_1 = 0.06 ~,~~~~ n = 3 ~,~~~~ \mu_0 = 1 ~,~~~~ \mu_1 = 0.8
  ~,~~~~ t = 1 ~.
 \end{align}
 \centerline{
 \pb{145mm}{
 \centerline{
   \includegraphics[scale=0.35]{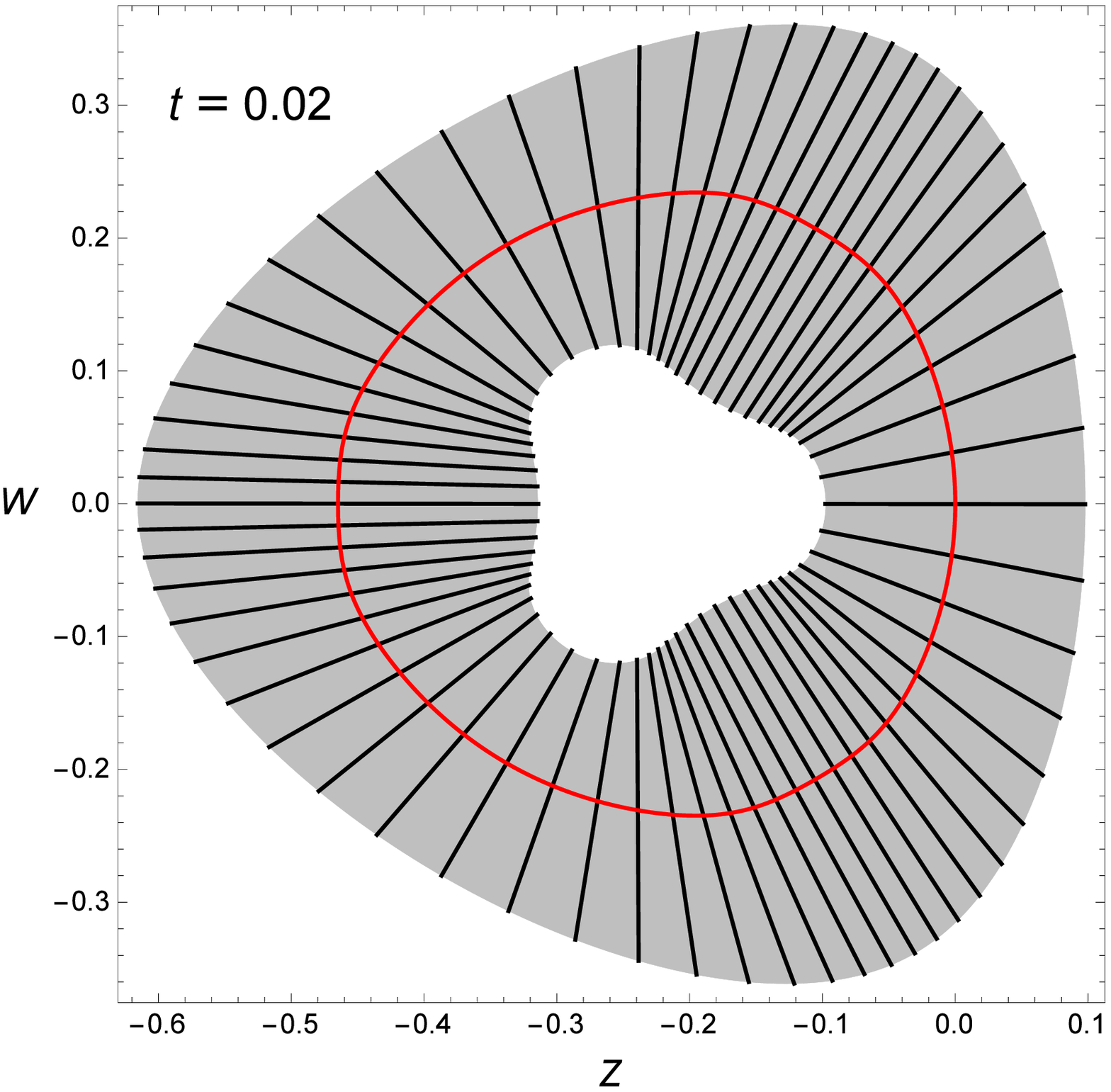}
   ~~~~
   \includegraphics[scale=0.34]{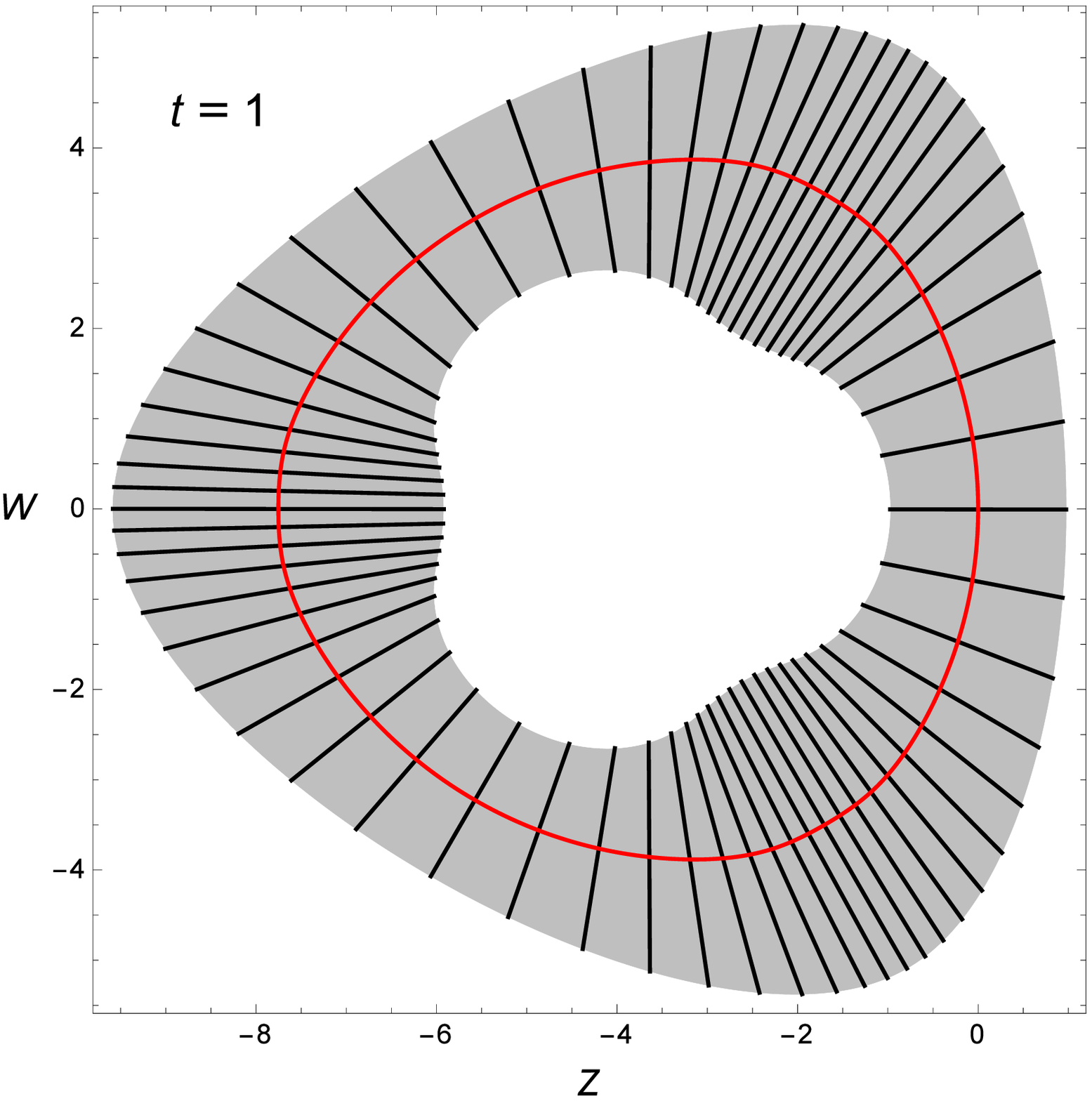}
 }
 \centerline{
   \includegraphics[scale=0.35]{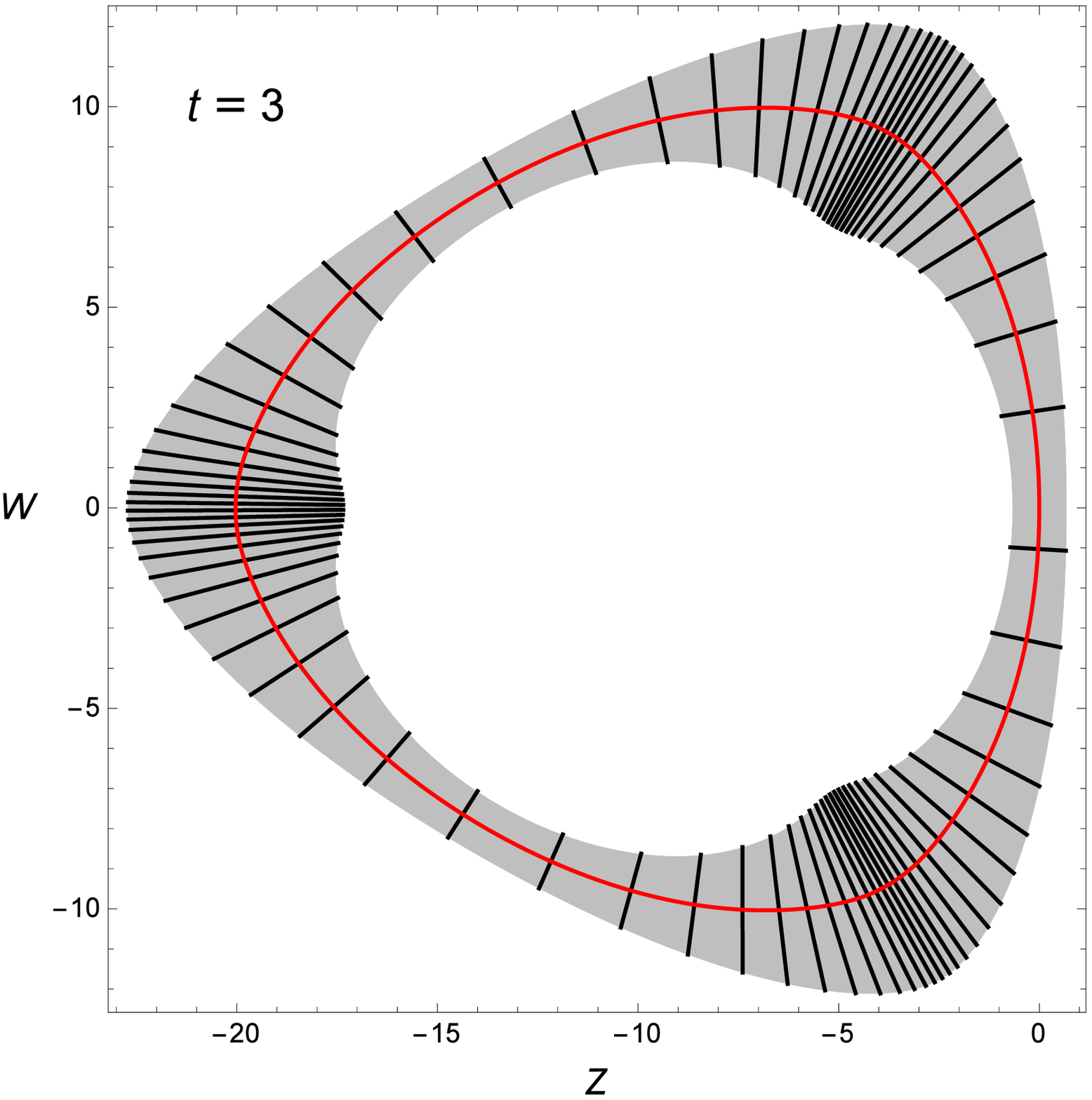}
   ~~~~
   \includegraphics[scale=0.365]{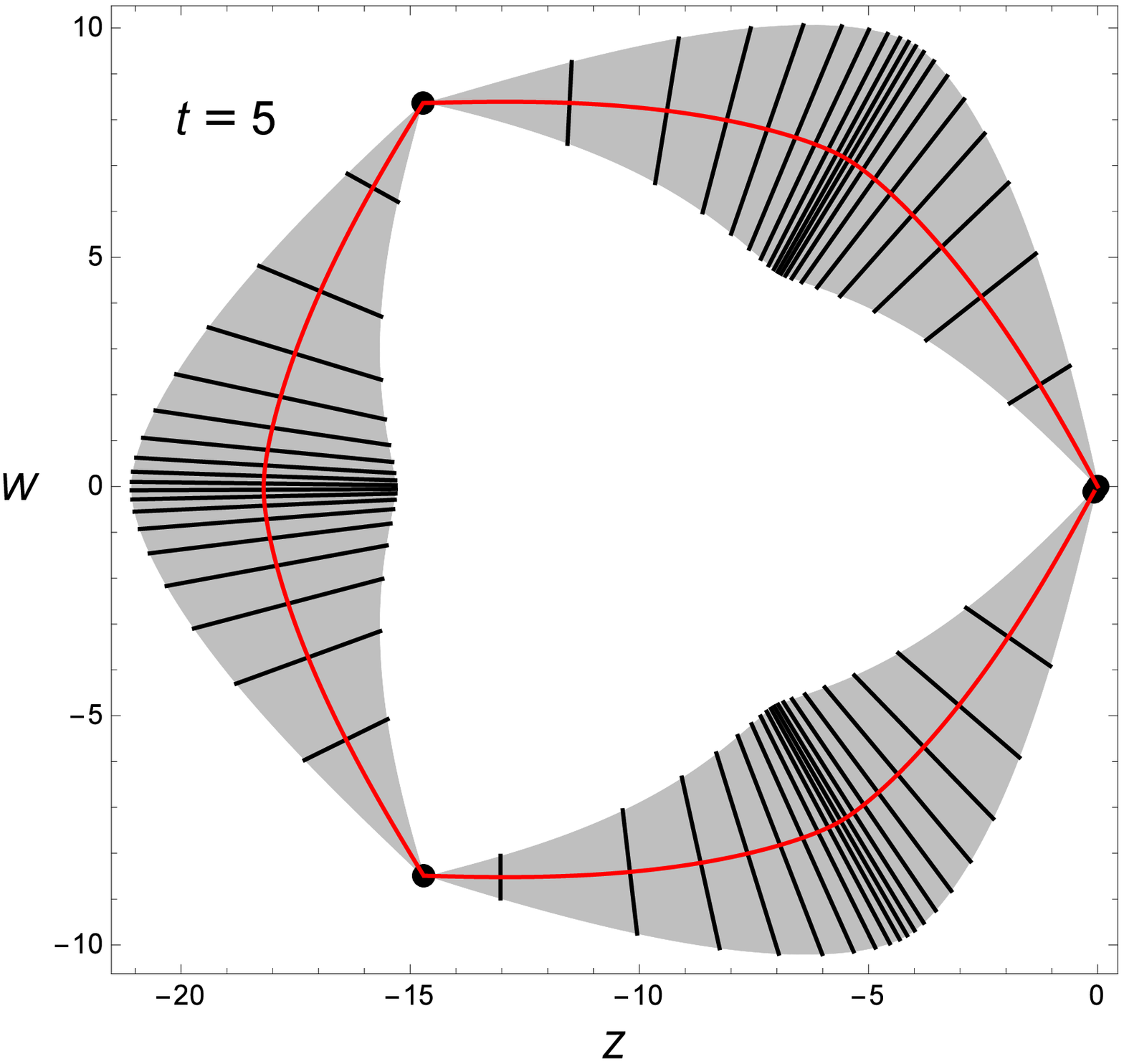}
 }
 \refstepcounter{figure}
 {\small Fig \arabic{figure}\showlabel{Md2Plt}.~~ An embedded Szekeres torus model with 3 lobes, showing just its intersection with the $(Z,W)$ plane.  The red curve is the path of the sphere centres, and each black line is the diameter of a 2-sphere at a particular $r$ value.  The arbitrary functions and parameter values of this model are listed in section \ref{Mdl2S}.  The times of the plots are $t = 0.02$, $t = 1$, $t = 3$, and $t = 5$.
 }}}

 What's interesting here is that there are parts where the tilt is changing more rapidly and parts where it's hardly changing.  Although the effect found in section \ref{CnLnCv}, that the dipole displacement and the tilt have opposing effects on the curvature of the path of centres near a spatial minimum, is evident here, in the models we tried, after shell crossings had been eliminated, the tilt curvature seems to dominate.

 \subsection{Model 3, $P$ Only}
 \showlabel{Mdl3P}

 For this model, the only non-sphericity function we vary is $P$.  By \cite{GeoHel17} we expect the line of centres to be bent, and by section \ref{SzRot}, we expect there to be shell rotation.  It is defined by
 \begin{align}
 \begin{aligned}
   M(r) & = M_0 + M_1 \big\{ 1 - \cos(n \mu r) \big\} ~, \\
   f(r) & = - 1 + f_1 \big\{ 1 - \cos(2 n \mu r) \big\} = - 1 + 2 f_1 \sin^2(2 n \mu r) ~, \\
   a(r) & = 0 ~, \\
   S(r) & = 1 ~,~~~~~~ Q(r) = 0 ~, \\
   \zeta & = \mu r ~, \\
   P'/S & = \zeta'/\alpha ~, \\
   \to~~~~~~~~ \alpha & = \frac{\sqrt{1 - 2 f_1 \sin^2(n \mu r)}\;}{\sqrt{2 f_1}\; \sin(n \mu r)} ~, \\
   P' & = \frac{\mu \sqrt{2 f_1}\; \sin(n \mu r)}{\sqrt{1 - 2 f_1 \sin(n \mu r)^2}\;} ~, \\
   P(r) & = \frac{1}{2 n} \ln \Bigg( \frac{(1 - 2 f_1 \{ 1 - 2 \cos^2(n \mu r) \} )}{2 \sqrt{2 f_1}} \\
   &~~~~~~~~~~~~~~ - \cos(n \mu r) \sqrt{1 - 2 f_1 \sin^2(n \mu r) }\; \Bigg) ~, \\
   M_0 = 0.75 ~,~~~~ M_1 & = 0.25 ~,~~~~ f_1 = 0.02 ~,~~~~ \mu = 1 ~,~~~~ n = 6 ~,~~~~ t = 1 ~.
 \end{aligned}
 \end{align}
 and is illustrated in fig \ref{Md3Plt}.
 \\
 \centerline{
 \pb{145mm}{
 \centerline{
   \includegraphics[scale=0.35]{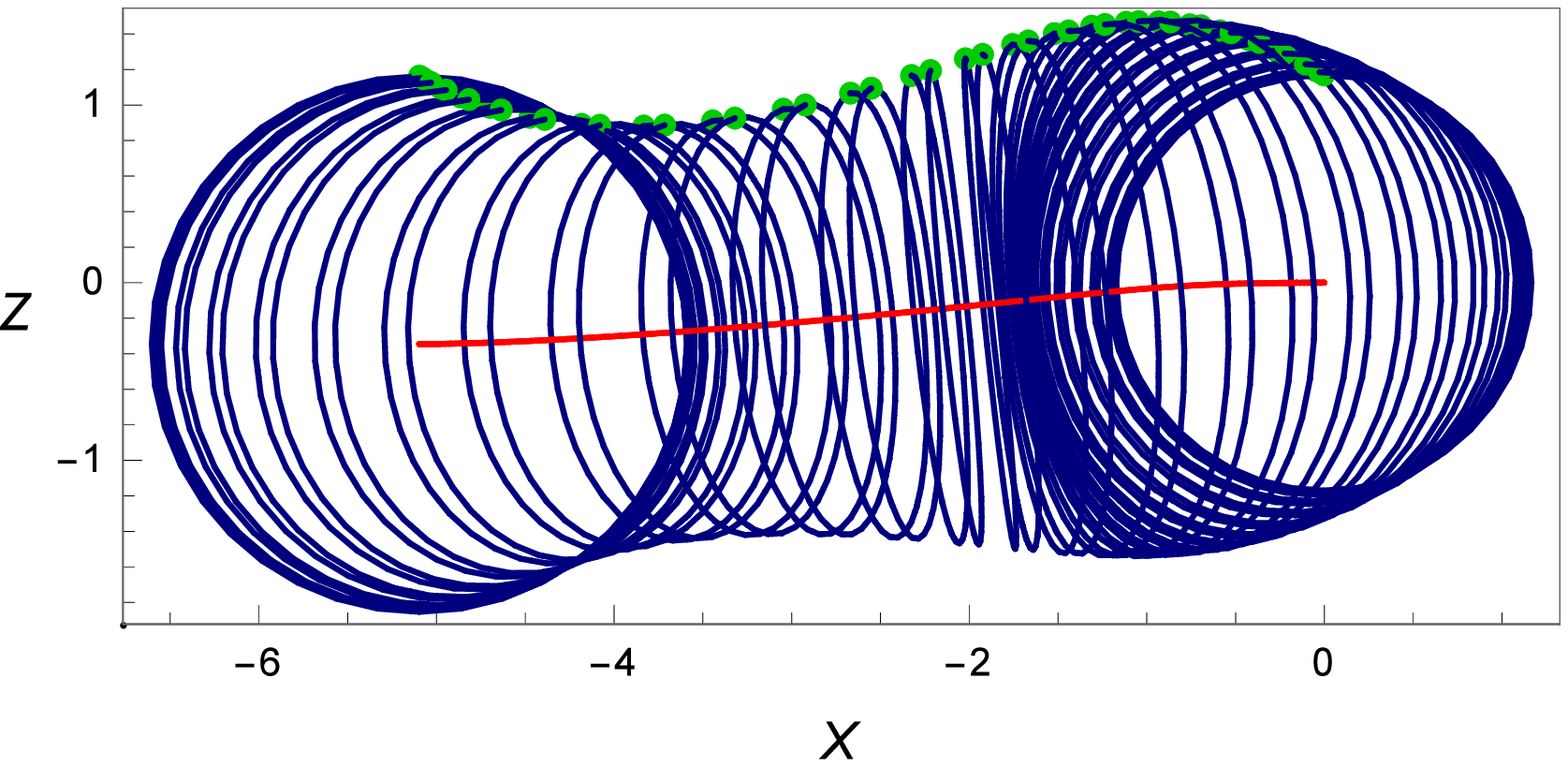}
 }
 \centerline{
   \includegraphics[scale=0.34]{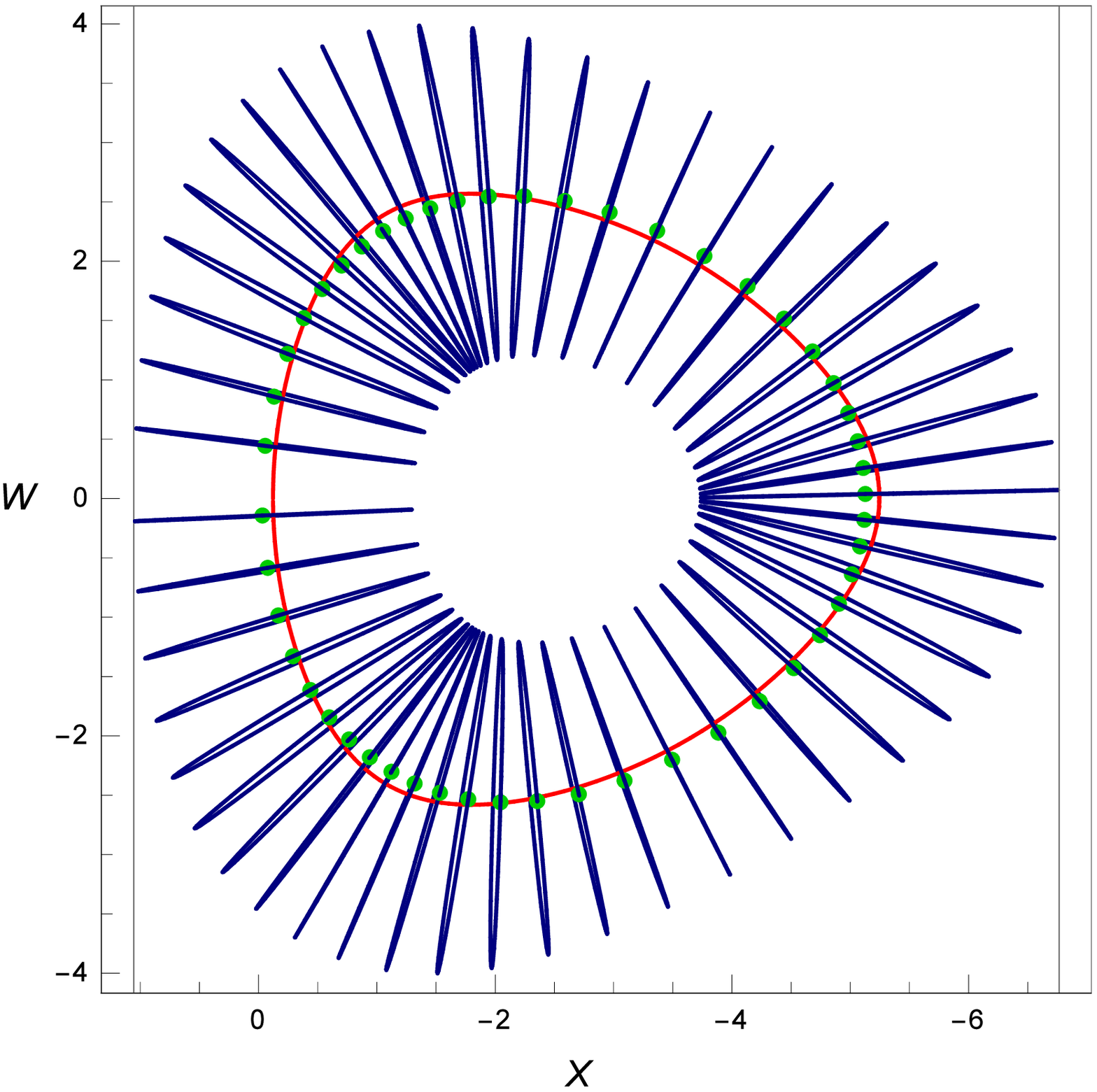}
   ~~~~
   \includegraphics[scale=0.365]{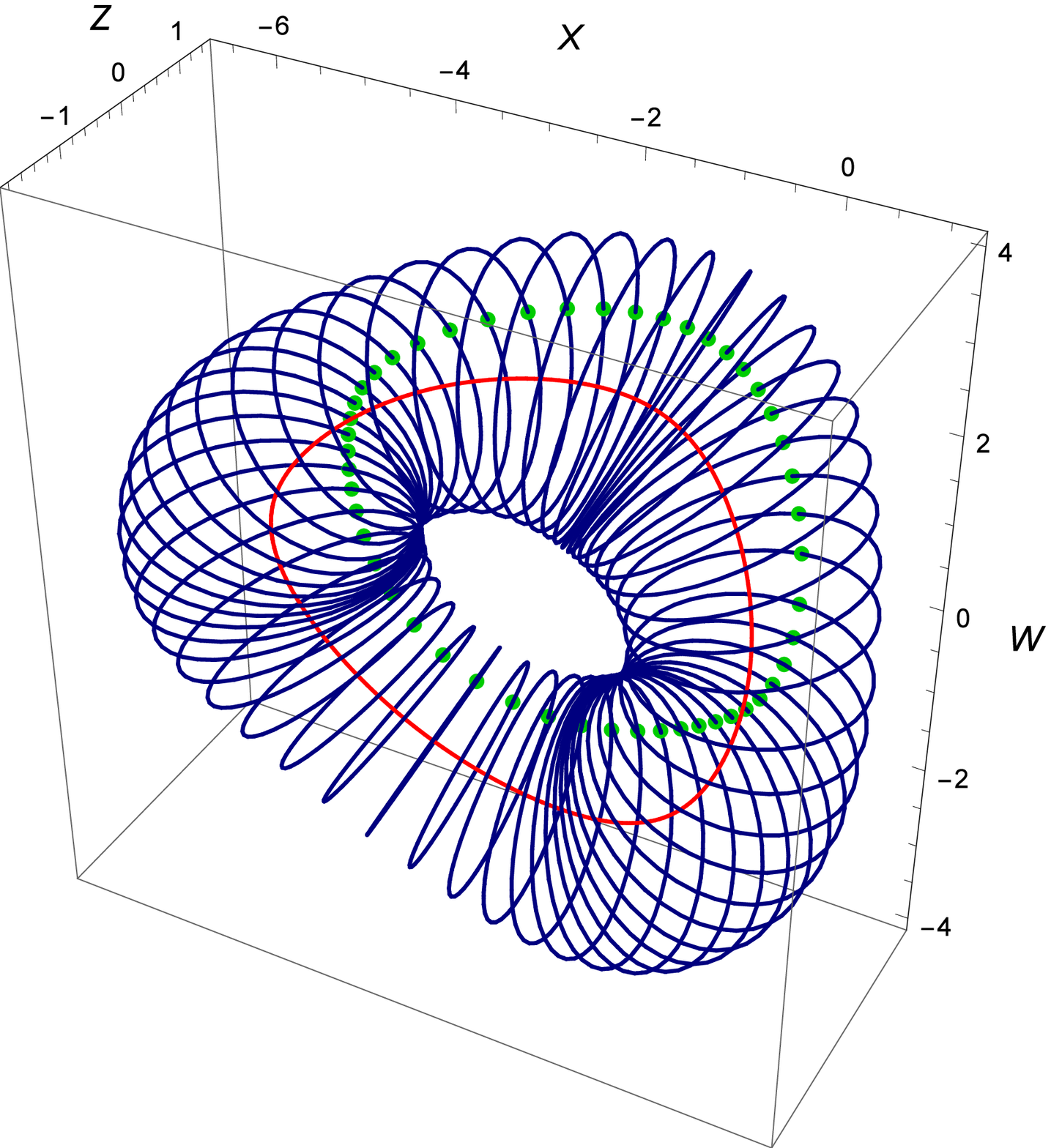}
 }
 \refstepcounter{figure}
 {\small Fig \arabic{figure}\showlabel{Md3Plt}.~~ Three views of a less symmetric embedded Szekeres torus model, in the $(X,Z,W)$ 3-space, at time $t = 1$.  The red curve is the path of the sphere centres, each blue circle is the outline of a 2-sphere at a particular $r$ value, and the green dots indicate the location of $\theta = 0$ on the 2-sphere; the rotation of the $(\theta,\phi)$ coordinates is not very large in this model.  The arbitrary functions and parameter values of this model are listed in section \ref{Mdl3P}.
 }}}

 Apart from the toroidal topology, the line of centres no longer lies in a 2-plane, and the direction of the `north pole' $\theta = 0$, rotates between shells; however these variations are fairly small.

 \subsection{Model 4, $P$ Only}
 \showlabel{Mdl4P}

 This model is not toroidal; it is a spatially closed model defined by
 \begin{align}
 \begin{aligned}
   M(r) & = \sin^3(\mu r) \big\{ M_0 + M_1 \sin(\mu r) \big\} ~, \\
   f(r) & = - \sin^2(\mu r) ~, \\
   a(r) & = \frac{- \pi M}{(-f)^{3/2}} ~, \\
   S(r) & = 1 ~,~~~~~~ Q(r) = 0 ~, \\
   P(r) & = \sin(\mu r) \big\{ P_0 + P_1 \sin(\mu r) \big\} ~, \\
   \to~~~~~~~~ \alpha & = \tan(\mu r) ~, \\
   M_0 = 1 ~,~~~~ M_1 & = 5 ~,~~~~ P_0 = 1 ~,~~~~ P_1 = 0.3 ~,~~~~ \mu = \pi ~,~~~~ \eta = 8 \pi / 5 ~.
 \end{aligned}
 \end{align}
 Fig \ref{M4Slc} shows 3 views of the resulting embedding.  The right hand view shows there is substantial rotation of the $\theta = 0$ direction.

 \centerline{
 \pb{145mm}{
 \centerline{
   \includegraphics[angle=-90,scale=0.4]{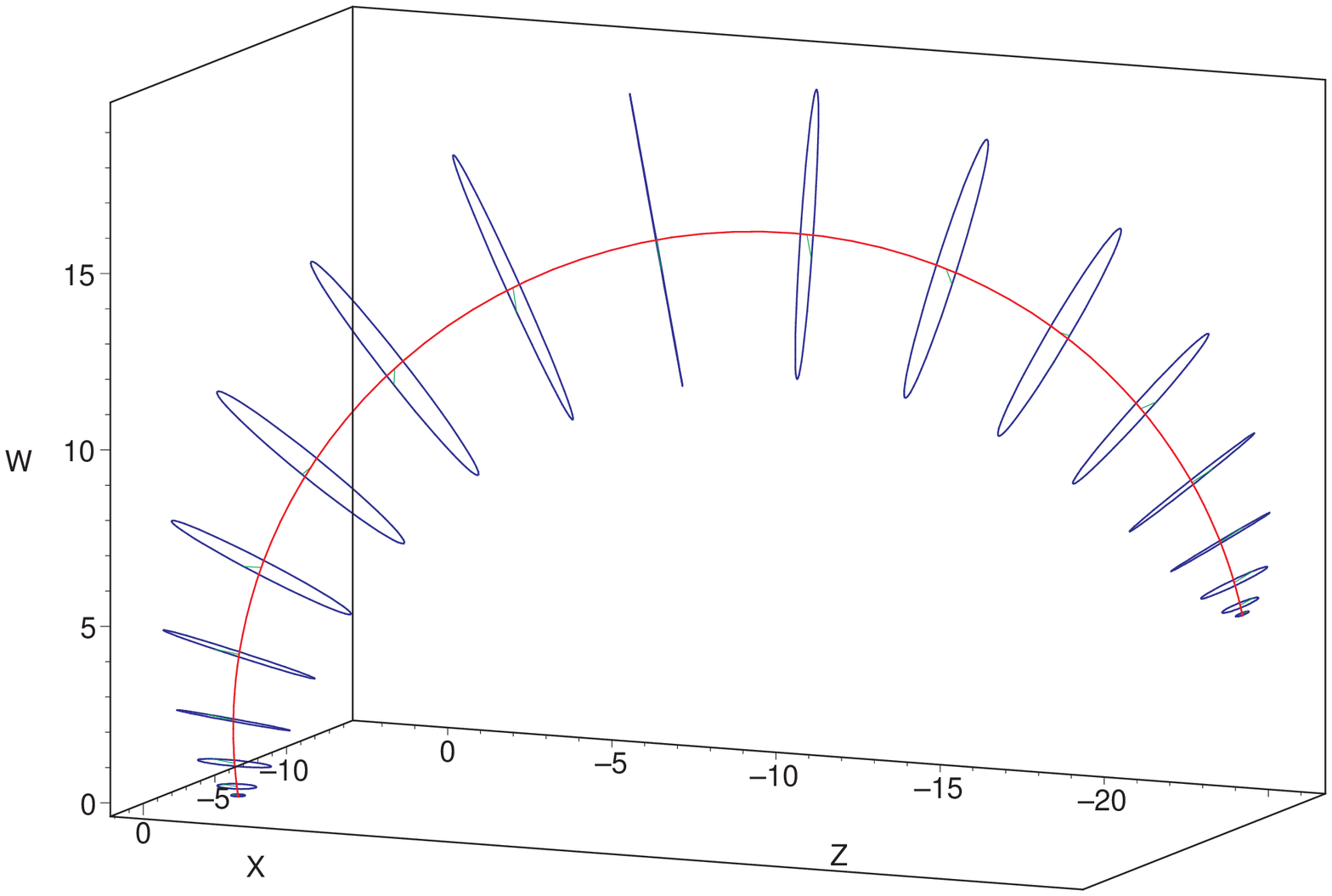}
 }
 \centerline{
   \includegraphics[angle=-90,scale=0.4]{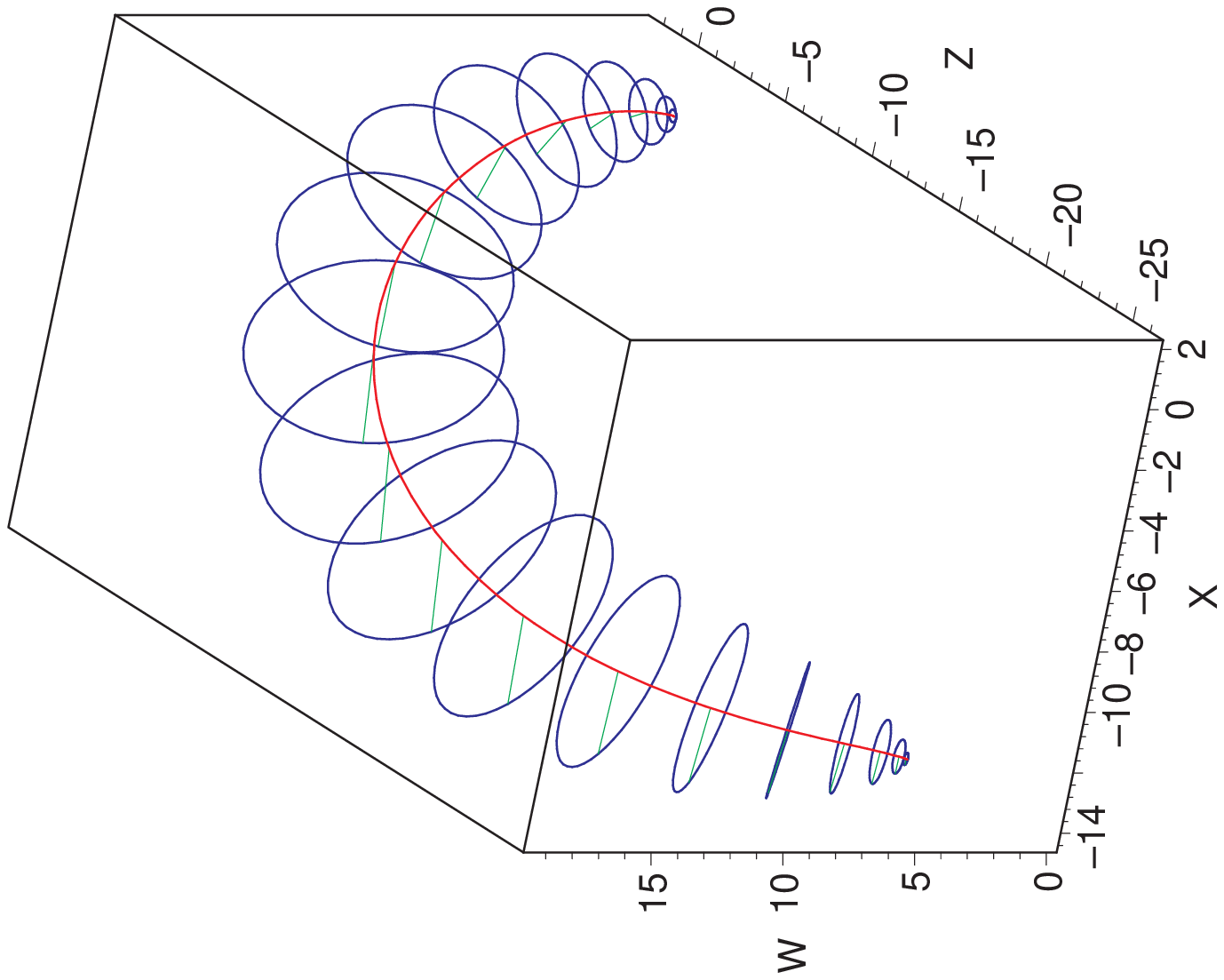}
   ~~~~
   \includegraphics[angle=-90,scale=0.4]{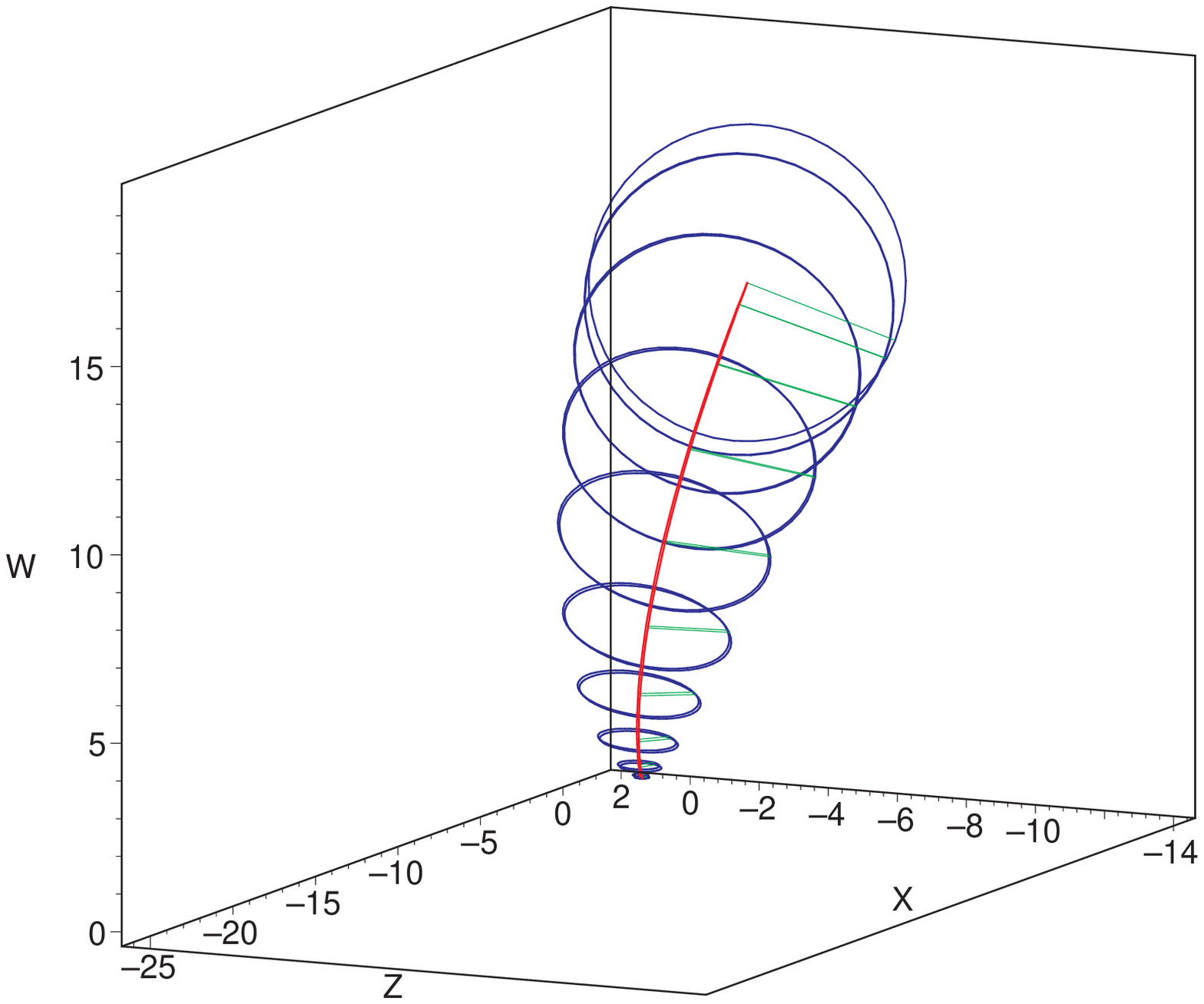}
 }
 \refstepcounter{figure}
 {\small Fig \arabic{figure}\showlabel{M4Slc}.~~ Three views of the embedded Szekeres model 4, at time $\eta = \pi$ which corresponds to $t = 0$.  The blue circles represent 2-spheres of constant $r$; tilting relative to the line of sight makes some appear elliptical.  The red curve is the centre path --- the locus of sphere centres, and the green lines show the direction of $\theta = 0$ for each plotted sphere/circle.  The $Y$ coordinate is suppressed, and the centre path lies in $Y = 0$.
 }}}

 \subsection{Model 5, RW as Szekeres}
 \showlabel{Mdl5P}

 The Szekeres models become homogeneous, despite the non-symmetric coordinates, if the LT functions $f$, $M$ \& $a$ take the RW form, regardless of the $S$, $P$ \& $Q$ functions%
 \footnote{\sf See \cite{GooWai82} around eq (2.24) and section 1.3.4 of \cite{Kras97}.}%
 .  Such a model is obtained with $M = M_0 \sin^3(\mu r)$, $f = - \sin^2(\mu r)$, $a = 0$, $S = 1$, $Q = 0$ and $P = P_0 \sin(\mu r) (P_0 + P_1 \sin(\mu r))$, along with parameter values $M_0 = 1$, $P_0 = 0.7$, $P_1 = 0.09$, $\mu = \pi$, and the result is shown in fig \ref{M5Slc}.

 \centerline{
 \pb{145mm}{
 \centerline{
   \includegraphics[angle=-90,scale=0.4]{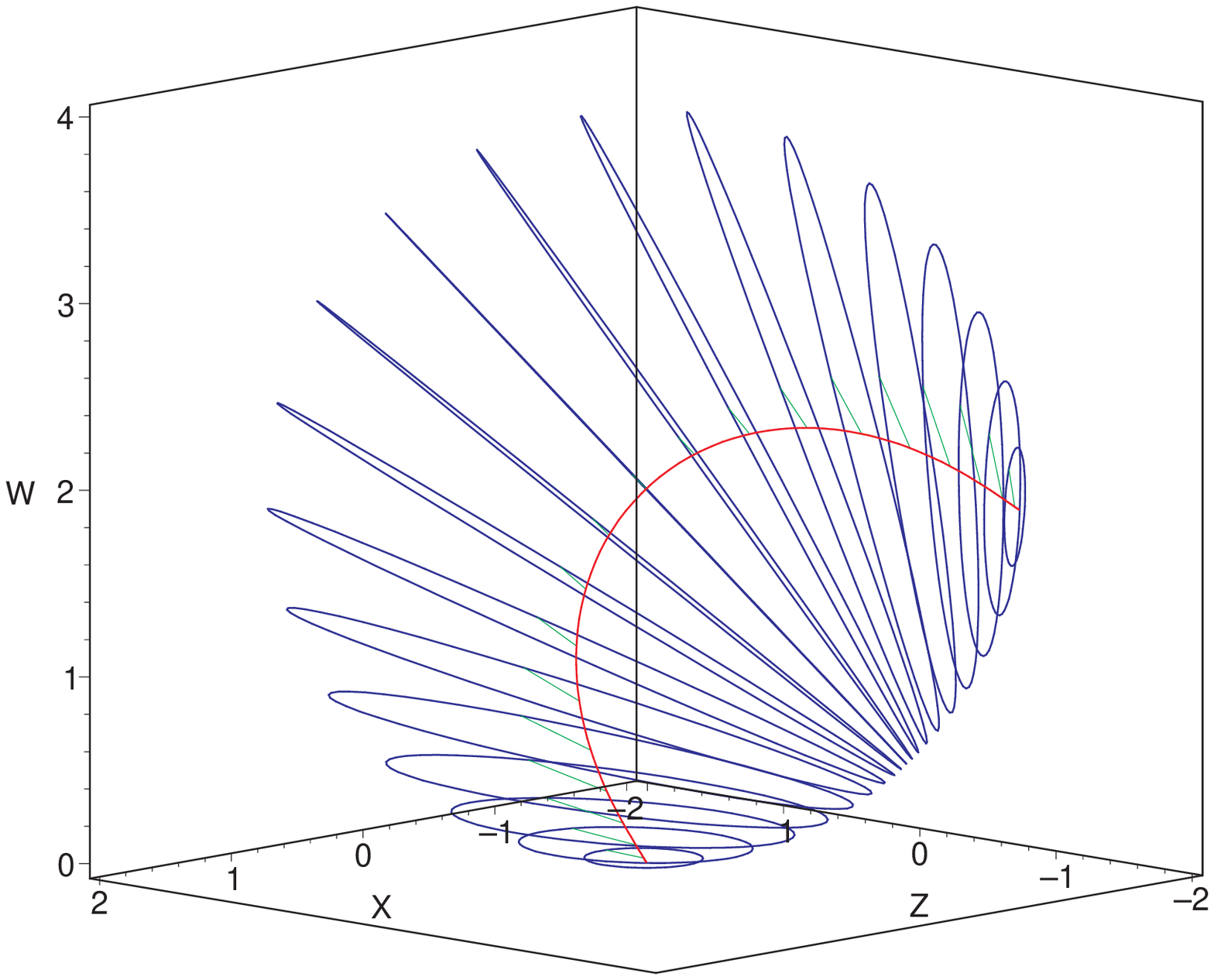}
 }
 \centerline{
   \includegraphics[angle=-90,scale=0.4]{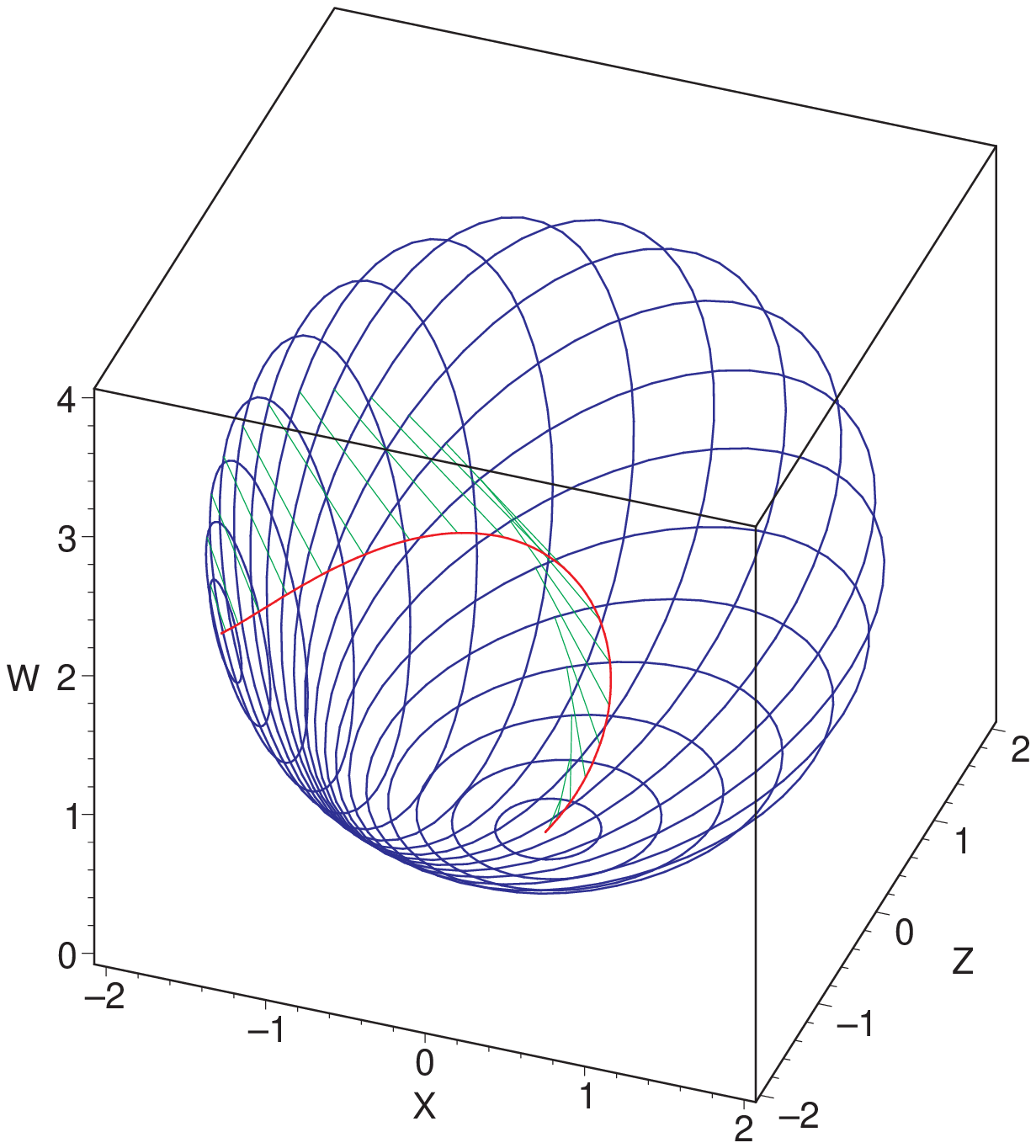}
   ~~~~
   \includegraphics[angle=-90,scale=0.4]{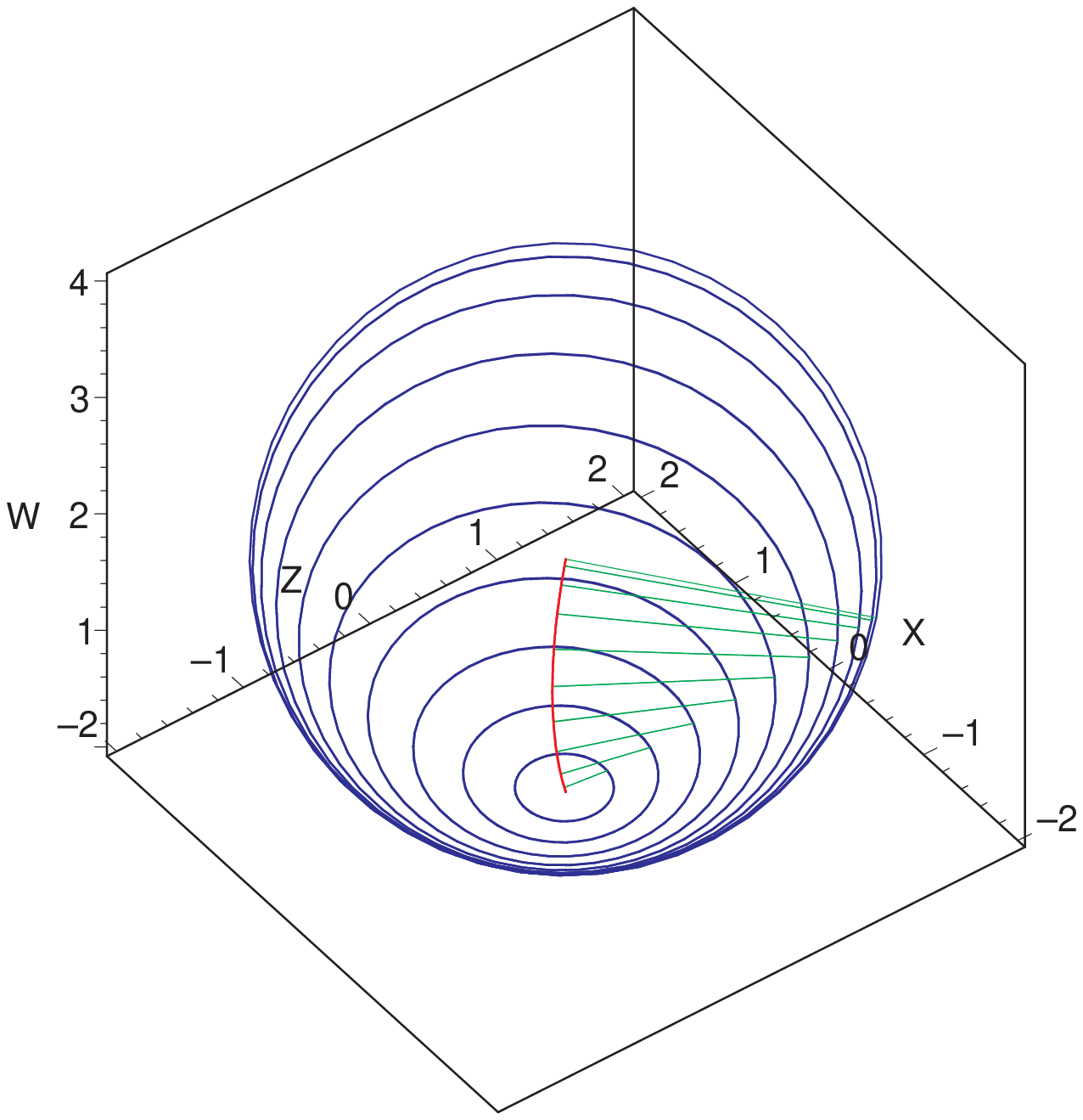}
 }
 \refstepcounter{figure}
 {\small Fig \arabic{figure}\showlabel{M5Slc}.~~ Three views of the embedded RW-Szekeres model 5, showing it is actually a 3-sphere in very non-symmetric coordinates.  The centre path (red line) is bent round, and the `north' (green lines) show significant variation.
 }}}

 Although the overall embedded shape is clearly a 3-sphere, the slicing is not only non-parallel, it also has shell rotation as shown by the variation of the `north' direction.

 \section{Conclusions}

 A correct understanding of the geometry of the Szekeres metric is important both for physical interpretation, when it is used for models of inhomogeneous gravitating structures, and also for a more accurate graphical depiction of those models.  Though the `non-concentric' property of constant $r$ shells was known from the start, it was only recently shown there is a hidden shell rotation effect, that appears when the usual angular coordinates are used.  Hitherto, it was tacitly assumed these coordinates had a constant orientation, as is the case for {\LT} (LT) models.

 In this paper we have shown that two independent results about Szekeres shell rotations are in full agreement.  In FR \cite{Hell17} it was shown, using the rotation rate of an orthonormal tetrad, that the $(\theta, \phi)$ coordinates of \er{ds2SzAng} do not in general retain any kind of constant orientation.  In \cite{BucSch13} the relative rotation of adjacent $(\theta, \phi)$ shells was stated, and in PG \cite{BucSch19} this was explained in several ways, notably an embedding that further introduced higher dimensional tilts.  The forms of the results in FR \& PG are sufficiently different that an alignment is called for.  We have reviewed the embedding of a $f < 0$ Szekeres model in flat 4-d Euclidean space, and discussed its visualisation.  It is striking that the embedding of a Szekeres model is locally the same to first order as that of the underlying LT model at the corresponding $r$ value.  By showing how to derive the FR results from the PG results, we have confirmed the reality of these rotations and tilts, and thereby their importance for graphing Szekeres slices.  Methods for this graphing are suggested in PG.

 To illustrate the higher dimensional tilts, we have constructed Szekeres models whose 3-spaces `naturally' have the topology of a torus, when embedded in a 4-d Euclidean flat space, without arbitrary identifications.  Explicit Szekeres models that are closed in the $r$ direction, with or without an arbitrary identification, had not been been previously considered.  Models 1 \& 2 had just $S$ varying, but nicely demonstrated the toroidal embedding as well as non-trivial radius and tilt structure.  Models 3 \& 4 had just $P$ varying, and additionally demonstrated a non-planar curve of shell centres.  Importantly, model 4 clearly exhibited the shell rotation described in PG.  Clearly, models in which all 3 of the Szekeres functions $S$, $P$ \& $Q$ vary could produce even more interesting embedded shapes.

 Some questions for future investigations are \\
 (i)  Can one find a different $(p, q)$ to $(\theta, \phi)$ transformation that incorporates the rotation, \er{ShellRot}, found by Buckley \& Schlegel?  Is it reasonably neat or too complicated?  Does the resulting metric look at all useful or  useable? \\
 (ii)  Can one apply or extend the embedding to the DKS-type (``$\beta' = 0$'') Szekeres models? \\
 (iii)  Is there a reasonably simple or elegant embedding of the $\epsilon \ne + 1$ Szekeres models? \\
 (iv)  Is it possible to create a quasi-spherical Szekeres model in which the shell rotation turns the `north pole' through 180 degrees?  Can such a model be given a toroidal topology, of any kind?  That would make the transformation to $(\theta, \phi)$ coordinates clearly inconsistent.

 \end{document}